\documentclass[prd,nofootinbib]{revtex4}
\pdfoutput=1

\usepackage{amssymb}
\usepackage{amsmath}
\usepackage{graphicx}
\usepackage{longtable}
\usepackage{verbatim}
\usepackage{amsfonts}
\usepackage{hyperref}

\newcommand{\matel}[3]{\langle #1|#2|#3\rangle} % Roman

\newcommand{\al}{\alpha}
\newcommand{\be}{\beta}
\newcommand{\ga}{\gamma}

\newcommand{\la}{\lambda}

\newcommand{\gpol}{\omega}
\newcommand{\HA}{H}
\newcommand{\vv}{\kappa}

\newcommand{\ee}{\end{equation}}
\newcommand{\bea}{\begin{eqnarray}}
\newcommand{\eea}{\end{eqnarray}}
\newcommand{\beq}{\begin{equation}}
\newcommand{\eeq}{\end{equation}}
\def\beqa{\begin{eqnarray}}
  \def\eeqa{\end{eqnarray}}
\newcommand{\bv}{\left(\begin{array}{c}}
\newcommand{\ev}{\end{array}\right)}

\def\lsim{\mathrel{\rlap{\lower4pt\hbox{\hskip1pt$\sim$}}
    \raise1pt\hbox{$<$}}}         %less than or approx. symbol
\def\gsim{\mathrel{\rlap{\lower4pt\hbox{\hskip1pt$\sim$}}
    \raise1pt\hbox{$>$}}}         %greater than or approx. symbol

\newcommand{\GeV}{\,{\rm GeV}}
  
\newcommand{\AAA}{ |A|^2}
\newcommand{\RRR}{R}
\newcommand{\III}{I}
\newcommand{\epq}{q^2_{\rm max}}

\newcommand{\pA}{q_A}
\newcommand{\pB}{q_B}
\newcommand{\pBa}{q_{B_1}}
\newcommand{\pBb}{q_{B_2}}

\newcommand{\pC}{q_C}
\newcommand{\pI}{q_r}
\newcommand{\laI}{\la_r}
\newcommand{\R}[1]{\text{\bf#1}}

\newcommand{\CPC}{C_{\rm +}} % C_+
\newcommand{\CPV}{C_{\rm -}}  % C_-
\newcommand{\CPCPV}{C_{\rm \pm}}  % C_\pm
\newcommand{\CPVPC}{C_{\rm\mp}}   % C_\mp

\def \thl {{\theta_\ell}}
\def \thK {{\theta_{K}}}

\begin{document}

\begin{flushright}
DO-TH 13/23 \\
QFET-2013-08 \\
Edinburgh /13/22 \\
CP3-Origins-2013-037 DNRF90 \\
DIAS-2013-37
\end{flushright}

\vspace*{-7mm}

\title{(A)symmetries of weak decays at and near the kinematic endpoint}

\author{Gudrun Hiller}
\affiliation{Institut f\"ur Physik, Technische Universit\"at Dortmund, D-44221 
Dortmund, Germany}
\author{Roman Zwicky}
\email{Roman.Zwicky@ed.ac.uk}
\affiliation{School of Physics \& Astronomy, University of Edinburgh, Edinburgh EH9 3JZ, Scotland}

\vspace*{1cm}

\begin{abstract}
At the kinematic endpoint  of zero recoil  physical momenta are parallel which leads
to symmetries in the decay distributions.  We implement this observation for decays
of the type $A \to (B_1 B_2) C$ by extending the helicity formalism to include an unphysical timelike polarisation. The symmetries of the helicity amplitudes are worked 
out for a generic dimension six Hamiltonian for a  $B \to V \ell \ell$ decay type. 
We obtain \emph{exact} predictions for angular observables, e.g.,~for  the fraction of longitudinally polarized vector mesons, $F_L = 1/3$, which may be used to guide experimental analyses.
We investigate the vicinity of the endpoint  through an expansion 
in the three momentum of the vector meson.  New physics can be searched for 
in the slope of the observables near the endpoint. 
Current experimental data on $B \to K^* \ell \ell$ decays are found to be in agreement
with our predictions within uncertainties.
Application to other semileptonic $B$ and $D$ decays, including  $B \to V \ell^+ \ell^-$, $V=K^*,\phi, \rho$ and $B \to V \ell \nu$, $V=\rho,D^*$ is straightforward. 
For hadronic modes of the types $B \to V p \bar p, V \Lambda \bar \Lambda, ..$ and $B \to V \pi  \pi, V \pi K, .. $ endpoint relations apply as long as they are not overwhelmed by sizeable final state interactions between the $V$ and the hadron pair.

\end{abstract}

\maketitle

\tableofcontents

%%%%%%%%%%%%%%%%%%%%

\section{Introduction}
\label{sec:HA}

\begin{figure}[ht]
\begin{center}
\includegraphics[width=0.4\textwidth]{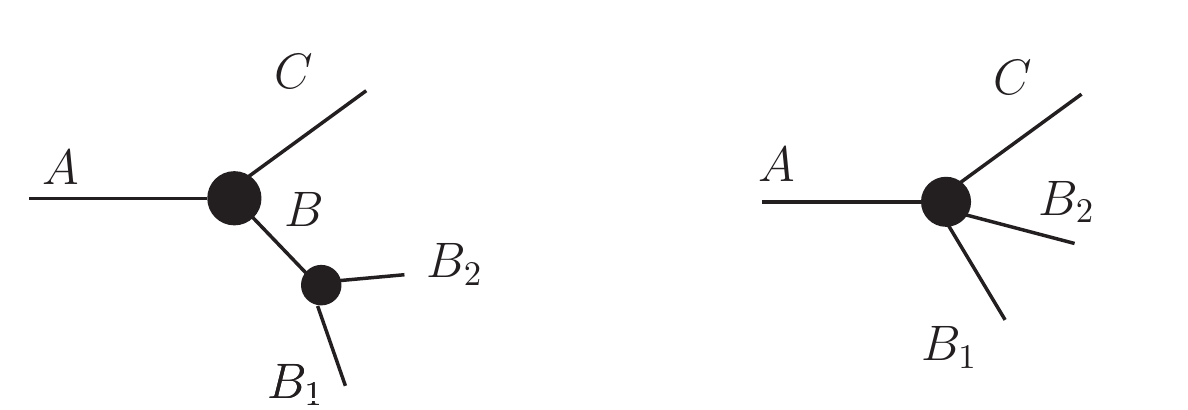}
\end{center}
\caption{\small Decay $A \to B(\to B_1 B_2) C$ with intermediate $B$-particle which is generically off-shell (left).
Decay $A \to ( B_1 B_2) C$ described by a point interaction (right). The endpoint configuration is described in the $A$-restframe by $\vec{q}_C =0$ and implies that $B_1$ and $B_2$ are back-to-back.  Possible decay products 
of the $C$-particle should have sufficiently suppressed interactions with the $(B_1 B_2)$-pair for the endpoint relation to remain visible in the decay.}
\label{fig:ABC}
\end{figure}

At the kinematic endpoint of a decay  the relevant spatial momenta are zero and, in the absence of  initial state polarization, there is no preferred axis.  This leads to isotropicity in certain observables. At the formal level it implies a reduction in the number of independent  Lorentz invariants.
We implement this idea for decays of the type $A \to  ( B_1 B_2)  C( \to C_1 C_2)$\footnote{The extension to  $A \to  ( B_1 B_2 .. )  C( \to C_1 C_2 .. )$-type decays is straightforward although the multi-body angular distributions will grow in complexity.}   for which 
the kinematic endpoint, in the $A$-restframe, is defined by $\vec{q}_C = 0$ and hence $\vec{q}_{B_1} + \vec{q}_{B_2} =0$. 
In the discussion it is implicit that the decay products of the $C$-particle do not interact significantly with the $(B_1 B_2)$-pair.
In the case where the $(B_1 B_2)$-pair  originates from a particle, denoted by  $B$  in 
Fig.~\ref{fig:ABC} (left),  the helicity formalism  \cite{JW59} (and \cite{Haber:1994pe} for a review) provides a powerful tool for the description 
of the kinematic endpoint \cite{RZ13}. The goal of this work is to extend this idea to the case where the decay is described, in part, by an  interaction of the type 
$H^{\rm eff} \propto A^\dagger C (B_1 B_2)$ as depicted in Fig.~\ref{fig:ABC} (right). 
This case is still amenable to a  helicity formalism provided one 
introduces an unphysical timelike polarisation direction, see section \ref{sec:extension}.

Concrete examples arise in flavor physics, where semileptonic
decays of  beauty and charm mesons are described by a $|\Delta F|=1$ effective Hamiltonian.  The latter contain terms  {\it e.g.,}  ${\cal{H}}^{\rm eff}_{b \to s \ell \ell}  \propto \bar s_L  \ga_\mu  b  \bar\ell\gamma^\mu \ga_5 \ell + ..\,$  originating from integrating out $Z$- and $W$-bosons 
which  \emph{cannot} be interpreted as originating from a sequential decay of the type $B \to K^* \gamma^*(\to \ell \ell)$.
 Predictions for angular observables at the endpoint are worked out for a large class of decays in section \ref{sec:app}; here, $B \to K^{*} \ell \ell$ serves as a template. 
 
It is noteworthy that the endpoint relations have to emerge in any consistent approximation that respects Lorentz invariance. 
The  low recoil operator product expansion (OPE) \cite{Grinstein:2004vb} for $B \to K^{*} \ell \ell$ is one such example. Here, 
after employing the improved Isgur-Wise relations \cite{Isgur:1990kf,Grinstein:2002cz}
 the endpoint symmetries are manifest \cite{Bobeth:2010wg}.  For recent applications, see   {\it e.g., }\cite{Bobeth:2012vn,Hambrock:2013zya}.
 We emphasise that  endpoint relations are not directly related to Isgur-Wise relations although  both emerge at small or zero velocities. Endpoint relations are of purely  kinematic nature and relate helicity amplitudes of different directions. The Isgur-Wise relations are based on QCD equation of motions and relate tensor and vector currents of the \emph{same} 
helicity direction. In the context of the aforementioned low recoil OPE they readily serve to
implement the heavy quark expansion on the level of matrix elements.

The exactness of the endpoint relations 
raises the question of the behaviour in the  vicinity of the endpoint-region. 
We investigate this question through an expansion in the  magnitude of the three-momentum $\vv = |\vec{\pC}|$  of the $C$-particle in the $A$-restframe.

The endpoint relations hold for the reconstructed decay in the endpoint configuration
where $\vec{q}_C = \vec{q}_{B_1} + \vec{q}_{B_2} =0$. Being based on kinematics 
the relations are independent of the production mechanism and prevail, as previously mentioned, provided the 
$C$-particle's decay products  do not significantly interact with the $(B_1 B_2)$-pair. 
For the $B \to K^* \ell \ell$-type decay this is the case as interactions are of the electroweak type and therefore negligible. For non-leptonic decays this is less evident 
as there can be inelasticities from rescattering into different final states. 
As discussed in section \ref{sec:range} this results in only a subset of events  remaining within the endpoint configuration. 
Furthermore, the relations do not apply in the case of missing momentum, {\it e.g.,} a soft photon being unnoticed in the detector. 
The process is then, however, a different one $A \to (B_1B_2) C \gamma$ at the price of shifting the
maximum  $(B_1 B_2)$-invariant mass to lower values depending on the photons' energy.

The paper is organised as follows: In section \ref{sec:extension} we discuss 
the helicity formalism in the effective Hamiltonian formulation.
In section \ref{sec:app} we work out the implications of our findings for 
angular decay observables for $B \to V \ell \ell $-type decays in the generic dimension six effective Hamiltonian. Sections \ref{sec:velocity} and \ref{sec:expansion} contain most of the phenomenological results.
We discuss $B \to K \ell \ell $-type decays as well as higher spin final states in section \ref{sec:otherthanV}.
To what extent endpoint relations remain valid for non-leptonic decays is investigated 
in section \ref{sec:range}.
In section \ref{sec:conclusions} we summarise.  Useful formulae for polarisation vectors
are compiled in appendix \ref{app:pol}.
The discussion of the  low recoil OPE for $B \to K^{(*)} \ell \ell $ \cite{Grinstein:2004vb} from the viewpoint of endpoint symmetries 
is deferred to appendix \ref{app:comment}. Parameterisations of helicity amplitudes
and considerations on asymptotic values of the fraction of longitudinal polarization are given in appendix \ref{app:q2expansion} and \ref{app:FLasymptotics}, respectively. Details on the tensor contributions  at the endpoint can be seen in appendix \ref{app:rJ}.

 \section{Endpoint relations for decays induced by effective Hamiltonians}
 \label{sec:extension}

 We discuss the reduction of  Lorentz-invariants  in terms of helicity amplitudes (HAs) 
 at the kinematic endpoint for $A \to (B_1 B_2) C$ type decays.
  Below we show using a generic language that such decays
 are amenable to a HA treatment provided one introduces a timelike polarisation reminiscent of covariant (Gupta-Bleuler) quantization of a  massive spin 1 boson.
 
 \subsection{The decay $A \to (B_1 B_2) C$ in terms of generalised helicity amplitudes}
 
The amplitude for a $A \to (B_1 B_2) C$ decay can be written in \emph{factorized} form 
 \begin{equation}
 \label{eq:like}
{\cal A}(A \to  (B_1 B_2) C ) =   (B_1 B_{2})_{\mu_1..\mu_X} \, C^{\mu_1 .. \mu_X} (\pA^2,\pB^2,\pC^2) \;.
\end{equation}
 The symbols $\pI,\laI$,  and $m_r$, $r =A,B,C$ denote the corresponding four momenta, helicities and masses, respectively, where  $\pB=  \pBa +\pBb$.
 In the notation above we have not imposed the on shell conditions $\pI^2 = m_r^2$. 
 The crucial point  in Eq.~(\ref{eq:like}) is that the momenta of $q_{B_{1,2}}$ do not enter the dynamics 
 as otherwise six invariants of a $1 \to 3$ decay would govern the form factor $C$ above.
 Contributions from 
 $A \to   (B^* \to B_1 B_2) C$, with $B$ being potentially off-shell, 
 are a special and simple case of Eq.~\eqref{eq:like}.
On a formal level interactions of the type in Eq.~\eqref{eq:like}  are still amenable to a helicity-like treatment provided one introduces a fourth (unphysical) helicity direction $\gpol(t,\pI) \propto (\pI)$  ($t$ stands for timelike). Some more details and remarks on conventions can be found  in appendix \ref{app:pol}. 
This allows for the following completeness relation 
 \begin{equation}
\label{eq:c-relation}
\sum_{\la,\la' \in \{t, \pm,0\}  }  \gpol^{\mu}(\la) \gpol^{*\nu}(\la')  G_{\la \la' }
= g^{\mu\nu} \;, \quad G_{\la \la' } = \text{diag}(1,-1,-1,-1) \;,
\end{equation}
where the first entry in $G_{\la \la' }$ refers to $\la = \la' = t$.
 Relation \eqref{eq:c-relation} can be inserted $X$-times to obtain a HA with $X$ helicity indices  in addition to the ones from the $A$ and $C$ particles.  In the notation of \eqref{eq:like} 
 the amplitude  may be written in terms of (generalised) helicity amplitude as follows,
 \begin{alignat}{2}
 \label{eq:HABC}
& {\cal A}(A \to  (B_1 B_2) C ) &\;=\;&    \HA_{ \la_{B_1} .. \la_{B_X},\la_C} 
(B_1 B_2)_{ \la_{B_1'} .. \la_{B_X'}}  G^{\la_{B_1} \la_{B_1'}} .. G^{\la_{B_X} \la_{B_X'}} \;,
  \nonumber \\[0.1cm]
  & (B_1 B_2)_{ \la_{B_1'} .. \la_{B_X'}}   &\;=\;&  (B_1 B_{2})^{\mu_1..\mu_X}\left[  \be_{\mu_1}(\la_{B_1}) 
 .. \be_{\mu_X}(\la_{B_X}) \right] \;,
  \nonumber \\[0.1cm]
&  \HA_{ \la_{B_1} .. \la_{B_X},\la_C} &\;=\;&  C^{\al \mu_1 ..\mu_X} \ga^*_\al(\la_C)  \left[ \be^*_{\mu_1}(\la_{B_1}) 
 .. \be^*_{\mu_X}(\la_{B_X}) \right]  \;,
 \end{alignat}
 with $J_A = 0$ and $J_C = 1$ for the sake of illustration. In the first equation we have implied the Einstein summation convention for the helicity indices.
Formally Eq.~\eqref{eq:HABC}  may be interpreted as a decay with  $X$ intermediate vector bosons which can take on physical as well as timelike polarisation directions. 
In the two-body decay $A(\la_A) \to B(\la_B) C(\la_C)$ where 
 the helicity quantization axis is taken to be the decay axis angular momentum conservation implies $\la_A = \la_B - \la_C$. More precisely it is the azimuthal symmetry around the decay axis that gives rise to this result. This can be used to define a 
 \emph{generalised helicity conservation (GHC) rule:} 
 Generically a polarisation vector $\zeta(\la,l)$  
transforms under a rotation around the $\vec{l}$-axis  by the azimuthal  angle $\phi$ as follows,
\begin{equation}
\zeta(t) \to \zeta(t)  \;, \quad \zeta(0) \to \zeta(0)  \;, \quad   \zeta(\pm 1) \to e^{\mp i \phi } \zeta(\pm 1) \; ,
\end{equation}
since $\sqrt{2} \vec{\zeta}_\pm = \vec{e}_1 \pm i \vec{e}_2$ where $\vec{e}_1$ and $\vec{e}_2$ are unit vectors orthogonal to $\vec{l}$.
The polarisation vectors  $\al(\la,\pA)$, $\be(\la,\pB)$ and $\ga(\la,\pC)^* =\ga(\bar \la,\pC)$, where
$\bar \la \equiv -\la$ throughout,
transform identically. 
Hence the GHC rule reads as follows:
\begin{equation}
\label{eq:rule}
\la_A =  \sum_{i=1}^X m( \la_{B_i})  +   \bar \la_C    \;, \quad m(t)=m(0) = 0 \;, \; m(\pm 1) = \pm 1   \; .
\end{equation}
 Below 
 we shall omit the function $m$ for brevity.   Since $\la_A$ is fixed by  Eq.~\eqref{eq:rule} it can be omitted from the labels in \eqref{eq:HABC}.
 Moreover, in the case where $\la_A =0$, as happens in the case of
 $B$ and $D$ meson decays, 
  $\la_C = \sum \la_{B_i}$ and it is 
 therefore possible and customary to omit $\la_C$ as well:
 \begin{equation}
 \label{eq:Hshort}
\la_A = 0: \quad  \HA_{ \la_{B_1} .. \la_{B_X},\la_C} \to  \HA_{ \la_{B_1} .. \la_{B_X}} \;.
 \end{equation}
 
In the remainder of this section we discuss the endpoint symmetries of the
HAs that enter the most generic dimension six  $b \to s \ell \ell$ effective Hamiltonian. The basis consists of $10$ operators  given in section \ref{sec:summary} together with a summary of endpoint relations.
We discuss the tensor operator, which is the most complicated case to analyze,  in the next section.

\subsection{The tensor operator transition for $B \to K^* \ell \ell$}
\label{sec:tensor}

We study tensor  operators in $b \to s \ell \ell$ transitions:
\begin{equation}
\label{eq:tensi}
H_{\rm eff} \propto  O_{\cal T}  + .. \;, \quad O_{\cal T}    \equiv \bar s_L \sigma^{\mu\nu} b  \bar\ell\sigma_{\mu\nu} \ell \;.
\end{equation}
In the notation of Eq.~\eqref{eq:like} $X=2$ and $(B_1 B_2)_{\mu_1 \mu_2} =  \bar\ell\sigma_{\mu\nu} (\gamma_5) \ell$.
The $\sigma$-Dirac structure corresponds to the antisymmetric  spin $1$ representation (in $SO(3)$ notation: 
$(\bf{3} \times \bf{3} )_{SO(3)} = \bf{3}_A  + ..$ where $A$ stands for antisymmetric and we denote the representations by $\bf{l}$ rather than its dimension $\text{dim}({\bf l}) = 2l +1$). We combine 
the two polarisation vectors  into an antisymmetric ``helicity"-tensor
\begin{equation}
\label{eq:anti1}
\beta_{\mu\nu} \equiv \beta_{\mu }(\la_{B_1}) \beta_{\nu }(\la_{B_2}) - \{\mu \leftrightarrow \nu \} \; ,
\end{equation}
not to be confused with a proper spin $2$ helicity tensor.
Using  GHC \eqref{eq:rule} we identify two classes of potentially non-vanishing HAs,
\begin{equation}
\label{eq:X12}
X_1  = \{ \HA_{t0}, \HA_{t+}, \HA_{t-} \} \;, \quad X_2  =  \{ \HA_{+-}, \HA_{+0}, \HA_{-0} \} \;.
\end{equation}
Above we have used the notation \eqref{eq:Hshort}. The helicity indices of the vector meson are $0,1$ and $-1$ in increasing order in the brackets.

For  $B \to K^* \ell\ell$ it is customary to denote by $p$ and $q$ the four momentum of
the vector meson and the lepton pair ($q_B = q$ and $q_C = p$ in the notation \eqref{eq:like}).
The HAs can be written in terms 
of Lorentz invariants $P_i$ as follows:
\begin{equation}
\label{eq:dec}
H_{\la_{B_1} ..\la_{B_X}} = \sum_{i=1}^N a_i(q^2) P_i( \beta(\la_{B_1}), .. ,\beta(\la_{B_X}), \ga(\la_C),  q , p)  \;,
\end{equation}
with $N$ finite at a certain order of effective Hamiltonian. 
We have omitted $p^2 = m_V^2$ and $(q+p)^2 = m_B^2$ from the form factor as
the corresponding particles are on shell.
At the endpoint we have got $p  \propto q  \propto (1,0,0,0) \equiv \gpol(t)$, leading
 to a reduction in the number of invariants  $N_e \leq N$ where 
the subscript $e$ stands for endpoint. In the case of sequential decays, that is without timelike polarisation $N_e \leq 1$, independent of the effective Hamiltonian \cite{RZ13}. We find $N=2$ for the tensor transition in two ways: {\it i)} writing down possible invariants and 
{\it ii)} using simple  arguments of representation theory with Clebsch Gordan coefficients (CGC).
\begin{itemize}
\item[{\it i)}] \emph{Through invariants:}  

At the endpoint $p \propto q$ implies $\gpol(\la) = \beta( \la,q) = \ga(\bar \la,p) $.
In attempting to build the number of invariants $N_e$ in  \eqref{eq:dec} we may therefore 
involve the following covariants $\gpol^*_{\mu\nu}(\la_{B_1} ,\la_{B_2})$ \eqref{eq:anti1}  $\gpol_\mu^*(\bar \la_C)$ and $\gpol_\mu^*(t)$.
Two types of invariants can be formed:
\begin{eqnarray}
\label{eq:P12}
(P_1)_{\la_{B_1 }\la_{B_2}} &\;=\;& \gpol^{*\mu\nu}( \la_{B_1 }, \la_{B_2}) \gpol_\mu^*(t) \gpol_{\nu}^*(\bar \la_C) \nonumber  \;, \\[0.1cm]
 \quad (P_2)_{\la_{B_1 }\la_{B_2}}   &\;=\;& \epsilon^{\mu\nu\rho\sigma} \gpol_{\mu\nu} ^*( \la_{B_1 }, \la_{B_2})\gpol^*_\rho(t) \gpol^*_{\sigma}(\bar \la_C)  \;.
\end{eqnarray}
As the notation suggests the identification with \eqref{eq:X12} is as follows:
$(P_i)_{\la_{B_1 }\la_{B_2}} \leftrightarrow X_i  = \{  \HA_{\la_{B_1 }\la_{B_2}} \} $ for $i =1,2$. 

\item[{\it ii)}] \emph{Through CGC:}
First, if only spatial indices are considered then $\gpol^*_{\mu \nu}$ corresponds to 
the antisymmetric vector representation $(\R{3} \times \R{3} )_{SO(3)} = \R{3}_A$ + ... 
Considering $\R{3}_A$ directly  corresponds to $P_2$ in \eqref{eq:P12}. 
The role of $\gpol^*(t)$ is to reduce the Lorentz Levi-Cevita tensor to the $SO(3)$ antisyemtric tensor $\epsilon_{ijk}$ (with $i,j,k=1..3$).
Second if $P_1$ is not to vanish, then $\gpol_\mu(t)$ can only be contracted with itself and does therefore not
bring in a new element. Hence the helicity $\la_C$ equals the helicity of $\la_{B_1}$ or $\la_{B_2}$.
In terms of CGCs:
\begin{alignat}{3}
&X_1: \;  \HA_{ t \la_{B_2} } &\sim& \; C^{011}_{0 \la_{B_2}   \bar  \la_{B_2}}  
\qquad \quad &   \leftrightarrow &  \quad   (\R{3}_{C} \times \R{3}_{B_1} ) = \R{1}  + .. \;,  
\nonumber \\[0.1cm]
\label{eq:CGC2}
&X_2: \;  \HA_{ \la_{B_1}  \la_{B_2} } &\sim& \; C^{011}_{0  \la_C   \bar \la_C } C^{111}_{\la_C \bar \la_{B_1} \bar \la_{B_2}}  &   \leftrightarrow & \quad  (\R{3}_{B_1} \times \R{3}_{B_2} )_{A} \times \R{3}_C = \R{1}   + ..  \;,
\end{alignat}
where we have used the notation  $\la_C = \la_{B_1} + \la_{B_2}$ in the second line 
of \eqref{eq:CGC2}.  The notation for the CGC is $C^{Jj_1j_2}_{M m_1m_2}$ for 
$j_1 \times j_2 = .. + J +..$ with $M = m_1 + m_2$.
In the equation above $\la_{B_1,B_2,C} = 0,\pm1$ (as opposed to $t$) 
only.
We have given the association with the $SO(3)$ Kronecker products and indicated  the
corresponding polarisation vectors  in the subscripts. The dots stand for higher representations
{\it i.e.} non-invariant objects which are  of no concern to  us.
\end{itemize} 
Using the parametrization given in appendix \ref{app:pol} for 
the polarisation tensors and tables for CGC {\it e.g.,} \cite{PDG}, the ratios of the HAs in both cases and both methods read\footnote{The results of the preprint v3 \cite{Bobeth:2012vn}
are different in signs because of different conventions for the polarisation vectors (see footnote \ref{foot:phases} for further remarks) 
and because of a mismatch of a factor $\sqrt{2}$ in
 $H_{t0}$ and $H_{+-}$. The final expressions, by which we mean the differential decay rate, yields results which are consistent for the purpose of this work.\label{foot:missmatch}}
\begin{alignat}{4}
& \HA_{t0} &\;:\;& \HA_{t+} &\;:\;&  \HA_{t-}  \;\;&=& \;-1 : 1 : 1 \;, \nonumber \\[0.1cm]
&  \HA_{+-} &\;:\;&  \HA_{+0} &\;:\;&  \HA_{-0}\;\; &=&\; -1 : 1 : 1  \;.
\end{alignat}
 Other operators contributing to \eqref{eq:like} can be analysed in analoguous manner with 
 either method.

 \subsection{Summary of endpoint relations for $B(D) \to V \ell \ell$}
 \label{sec:summary}
 
 Relations between HAs  of operators $B(D) \to V\ell\ell$ are summarised in this section.
 Endpoint relations for 
 $B(D) \to X_J \ell \ell$ for spin $J \neq 1$ are given in section \ref{sec:otherthanV}.
 The most generic dimension six operators are given by a total of $10$ operators 
 ($O_{\cal T}   =  1/2(O_T + O_{T_5})$ in the notation of \cite{Bobeth:2012vn}):
 \begin{eqnarray}
 \label{eq:operators}
 O_{S(P)} = \bar s_L   b \,  \bar\ell (\gamma_5) \ell\;,  \quad 
 O_{V(A)} = \bar s_L  \gamma^\mu b  \, \bar\ell \gamma_\mu   (\gamma_5) \ell\;, \quad 
 O_{\cal T} = \bar s_L \sigma^{\mu\nu}  b \,  \bar\ell\sigma_{\mu\nu} \ell  \;, \quad  O' = O|_{s_L \to s_R} \;.
 \end{eqnarray}
 We have assumed a $b \to s$ transition for the sake of explicitness. 
 The corresponding Wilson coefficients carry the same sub and superscripts, {\it e.g.,} $H_{\rm eff} \propto C^{'}_P \bar s_R   b \,  \bar\ell \ga_5 \ell+ ..$. Note, the notation $O_{V(A)} \propto O_{9(10)}$ is frequently used in the literature.  
 The basis \eqref{eq:operators} involves \emph{scalars}, \emph{vectors} and \emph{tensors} which correspond 
 to no, one and two-antisymmetric Lorentz indices. Non-vanishing HAs can also arise when operators with two-symmetric Lorentz indices \cite{HoZ13} are considered but 
 the latter are of dimension eight and are therefore expected to 
 be suppressed in generic models of new physics.
The endpoint relation of $O_{\cal T}$ were analysed  in the previous section.  
 Armed with this knowledge  and  appendix \ref{app:pol}, the relations of the scalar and vector operators are easily obtained:
\begin{itemize}
\item  \emph{scalars:} 
 at the endpoint the covariant objects are 
 $\gpol_\mu^*(\la_C)$ and $\gpol_\mu^*(t)$, which can be combined 
 into a scalar product which vanishes. Hence scalars do not contribute at the endpoint.  
 \item  \emph{vectors:} the covariant objects
 are $\gpol_\mu^*(\la_C)$,  $\gpol_\mu^*(t)$ and $\gpol_\mu^*(\bar \la_{B_1})$. 
 The only non-vanishing invariant at the endpoint is 
$ \gpol^*_\mu(\la_C)  \gpol^{*\mu}(\bar \la_{B_1}) = -(-1)^{\la_C} \delta_{\la_C \la_{B_1}}$, see
\eqref{eq:SP}. 
\end{itemize}
We summarise the endpoint relations\footnote{ \label{foot:phases} Note that  polarisation vectors are frequently  chosen 
such that $ H_0 =  H_1 = H_{\bar 1}$ and therefore $H_\parallel =  \sqrt{2} H_0$. We shall not do so as we stick to 
the Jacob-Wick and Condon-Shortley conventions of the second helicity state and CGC. Of course these conventions drop out in physical observables.}, valid at $q^2 = \epq = (m_B\!-\!m_V)^2$, as follows:
\begin{alignat}{3}
\label{eq:summary}
&\text{scalars \Big($O_{S,P}^{(')}$\Big):} \quad & & H = 0 \;,  & &    \nonumber \\[0.1cm]
&\text{vectors \Big($O_{V,A}^{(')}$\Big):}            & &H_{0} = - H_{+} = -H_{-}  \;, 
\qquad \quad & & \Big[ H_\parallel = - \sqrt{2} H_0 \;, H_\perp = 0 \Big] \;, \nonumber  \\[0.1cm] 
  & & &  H_{t} = 0  \;,  & &  \nonumber \\[0.1cm]  
&\text{tensors \Big($O_{\cal T}^{(')}$\Big):}   & & \HA_{+-} =-  \HA_{+0}=- \HA_{-0} \;, & & \Big[ H^{{\cal T}}_\parallel = - \sqrt{2} H^{{\cal T}}_0 \;, H^{{\cal T}}_\perp = 0 \Big] \;,\nonumber \\[0.1cm]  
& & &  \HA_{t0} = -  \HA_{t+}  = -   \HA_{t-} & & \Big[ H^{{\cal T}_t}_\parallel = - \sqrt{2} H^{{\cal T}_t}_0 \;, H^{{\cal T}_t}_\perp = 0 \Big] \;, 
\end{alignat}
where we have given the equivalent relations in the transversity basis, defined in the next section, in square brackets. 
The relevance of $O$ versus $O'$ in terms of selection rules of powers of the three momentum $\vv$ is discussed in 
the next section as well. The polarisation of the leptons, {\it e.g.,} $O_{S(V)}$ vs $O_{P(A)}$ is immaterial as the relations \eqref{eq:summary} describe  the $B \to V$ transition and 
do not specify the leptons in any way. 
Considering them can though provide additional information. 
For instance, $H_t$ vanishes or is proportional to a $m_\ell O_{P}$-contribution
when $H_t$ is contracted with a vector or axial lepton bilinear, respectively,
after using the equations of motion.
One might therefore write $ m_\ell H_t \to H^{\rm eff}_t$ and absorb  it into $O_P$ and vice versa.  We note that $H_t, H =0$ is consistent with the latter statement.

\subsection{Parity selection rules the transversity  amplitudes}
\label{sec:selection}

At the endpoint there are stringent selection rules on the HAs 
\cite{RZ13}. Following the same line of arguments we extend some of those results to include timelike polarisation
and obtain 
\begin{equation}
\label{eq:speed}
\text{HA} \propto \CPC \vv^{n} +  \CPV \vv^{n \pm 1} \;,  \quad \CPCPV \equiv C\pm C' \;,
\end{equation}
where $\vv$ is the absolute value of the three momentum of the $C$-particle in the A restframe.
It is proportional 
to the K\"all\'en-function $\lambda$:
\begin{equation}
\label{eq:speedD}
\vv = |\vec{\pB}| = |\vec{\pC}| =  \sqrt{ \frac{\la(\pA^2,\pB^2,\pC^2)}{4 \pA^2}}   \;,  \qquad      \la(\pA^2,\pB^2,\pC^2) \equiv (\pA^2 - (\pB+\pC)^2)(\pA^2 - (\pB-\pC)^2) \;,
\end{equation}
where $\pI \equiv \sqrt{\pI^2}$. The Wilson coefficients $C$ and $C'$ correspond to left handed and right handed $s$ quarks in 
  the transition operator, respectively \eqref{eq:operators}.

In general if parity is conserved then the S-matrix and therefore the amplitude
is an even(odd) power in the external momenta if 
the product of the internal parities of the initial and final state particles are 
$1(-1)$  \cite{Weinberg:1995mt}.
 It is therefore advantageous to consider HAs of definite parity (transversity basis):
\begin{alignat}{4}
&  H_{\parallel(\perp)} &\;\equiv\;& \frac{1}{\sqrt{2}}(H_+ \pm H_{-})  \;, & & & &   \nonumber \\[0.1cm]
& H_{\parallel(\perp)}^{{\cal T}_t} &\;\equiv\;& \frac{1}{\sqrt{2}}( H_{t+} \pm H_{ t-}) \;, \quad & &   
H_{0}^{{\cal T}_t} &\;\equiv\;& H_{t0}  \nonumber \\[0.1cm]
&  H_{\parallel(\perp)}^{{\cal T}} &\;\equiv\;& \frac{1}{\sqrt{2}}( H_{+0} \pm H_{ -0} )  \;, \quad & &   
H_{0}^{{\cal T}} &\;\equiv\;& H_{+-} \;.
\end{alignat}
To assess a definite parity, consider 
the HA for the $B(0^-) \to V(1^-)$ transition, which can be written as 
\begin{equation}
H_{\la_{B_1} .. \la_{B_X}} \propto \matel{V }{\bar b \Gamma_{\mu_1 ..\mu_X} s }{B} \beta^{\mu_1}(\la_{B_1}).. \beta^{\mu_X}(\la_{B_X} )  \;.
\end{equation}
The internal parity is 
$\eta = \eta_{B(0^-)} \eta_{V(1^-)} = (-1)(-1) = 1$. 
The parity of the transition is given by the parity of the  Dirac bilinear contracted with the polarisation vectors.  Concerning the latter one
needs to take  into consideration that $\beta_\mu(t)$ and  $\beta_\mu(i), \, (i = 0,\pm)$ transform as 
vectors $J^P = 1^-$ and pseudo-vectors $J^P = 1^+$ respectively.  
Consequences of parity covariance are summarised for $B(0^-) \to V(1^-) \ell \ell$ in table \ref{tab:selection}. The dots stand for corrections which are of \emph{relative} order 
${\cal O}(\vv^2)$ by virtue of parity and analyticity.
\begin{table}[h]
\begin{center}
\begin{tabular}{l || c ||  c |  c | c || c  | c | c  | c}
HA & $H$ & $H_t$  & $H_{0,\parallel}$ & $H_{\perp}$  & $H_{0,\parallel}^{{\cal T}_t}$ & $H_{\perp}^{{\cal T}_t}$  & $H_{0,\parallel}^{\cal T}$ & $H_{\perp}^{\cal T}$      \\[0.1cm] \hline 
WCC            & $\CPV$ & $\CPV$ & $\CPV$       &$\CPC$  &$\CPV$      &$\CPC$      &$\CPC$ &$\CPV$    \\[0.1cm]
$\propto \vv^n$& $\vv + ..$  & $\vv + ..$ &$\vv^0 + ..$ & $\vv + ..$ & $\vv^0 + ..$ & $\vv + ..$ & $\vv^0 + ..$ & $\vv + ..$
\end{tabular}
\end{center}  
\caption{\small Dependence of the HAs on the Wilson coefficient combination 
(WCC) for $B(0^-) \to V(1^-) \ell \ell$  decays.
The combination $\CPCPV \equiv C\pm C'$ corresponds to a parity conserving  (parity violating) coupling for the $b \to s$ transition. 
Together with the rule \eqref{eq:speed} and the discussion in the text this implies 
the power-behaviour in the three-momentum $\vv$ as indicated in the last row. 
For particles with opposite internal parities $\eta = -1$
but identical spin assignment, 
the substitutions $\CPCPV \to \CPVPC$ are the only requisite change to
this table.
An example is given by the decay $B(0^-) \to   K_1(1270)(1^+) \ell\ell$. }
\label{tab:selection}
\end{table}

We note that exchanging a spacelike with a timelike polarisation index selects 
the opposite chirality combination of Wilson coefficients in agreement with the transformation properties mentioned earlier. 
Table \ref{tab:selection} is consistent with the literature {\it e.g.} \cite{Bobeth:2012vn}. We would like to add that it equally applies for radially excited mesons. 
The rules  are easily adapted to different internal parities as described in the caption of the table.

\section{Semileptonic $B$ and $D$ decays into vector mesons}
\label{sec:app}

The endpoint relations \eqref{eq:summary} are of direct relevance for observables 
as the differential decay rate  is proportional to 
two powers of the HAs. Below we continue to use $\bar B \to \bar K^*(\to \bar K \pi)  \ell^+ \ell^-$ as a template for $S \to V \ell \nu$, $S \to V \ell^+  \ell^-$ and  $S \to V \nu \bar \nu$, where $S=B_{(s,c)},D_{(s)}$, $\ell=e,\mu, \tau$, 
and $V$ denotes a vector meson generally observed through the decay into two
pseudoscalar ones, see section \ref{sec:other}.
Applications to non-vector modes are discussed in section \ref{sec:otherthanV}.

The starting point for phenomenological implications is the $\bar B \to \bar K^*(\to \bar K \pi)  \ell^+ \ell^-$  angular distribution \cite{Kruger:1999xa}\footnote{We have taken out the overall phase space factor $\vv \vv_\ell$, which drops out in ratios. 
That is to say the decay rate is schematically given as $d\Gamma \propto \vv \vv_\ell |\text{HA}|^2 d (\text{phase space})$. \label{foot:root}} for 
the operator basis \eqref{eq:operators}:
\begin{eqnarray}
\nonumber
  \frac{8 \pi}{3 \vv \vv_\ell}  \frac{d^4 \Gamma}{d q^2\, d\!\cos\thl\, d\!\cos\thK\, d\phi} & = &
     (J_{1s} + J_{2s} \cos\!2\thl + J_{6s} \cos\thl) \sin^2\!\thK
\\ \nonumber
   &  + &(J_{1c} + J_{2c} \cos\!2\thl + J_{6c} \cos\thl) \cos^2\!\thK  
\\ \nonumber
   &  +& (J_3 \cos 2\phi + J_9 \sin 2\phi) \sin^2\!\thK \sin^2\!\thl
\\ \nonumber
   & +& (J_4 \cos\phi + J_8  \sin\phi) \sin 2\thK \sin 2\thl 
\\
   & + & (J_5 \cos\phi  + J_7 \sin\phi ) \sin 2\thK \sin\thl \, ,
\label{eq:anganal}
\end{eqnarray}
where $\theta_\ell$ denotes the angle between the $\ell^-$ and
$\bar{B}$ in the $(\ell^+\ell^-)$ center of mass system (cms), 
$\thK$ the angle between $K^-$ and $\bar{B}$ in the $(K^-\pi^+)$ cms and 
$\phi$ the angle between the two decay planes spanned by the 3-momenta of the
$(K^-\pi^+)$- and $(\ell^+\ell^-)$-systems, respectively. The variable $q^2$ denotes the
invariant mass-squared of the dileptons.
  The angular coefficients $J_i=J_i(q^2)$ expressed in terms of HAs    in the most general dimension six operator basis including finite lepton masses is given in Ref.~\cite{Bobeth:2012vn}.
  The phase space factor $\vv \vv_\ell$ is defined in Eq.~(\ref{eq:norm}).
   
From table \ref{tab:selection} we infer that there are four non-vanishing HAs, $H^{V,A}$ and $H^{{\cal T},{\cal T}_t}$ of the vector and tensor type. Since there are 
 a total of twelve angular functions $J_{ix}$, we therefore expect 
 at least 8 relations among them. In fact we find that there are ten relations:
\begin{alignat}{4} \label{eq:smaxlimitJi}
&  J_{2s}(q^2_{\rm max}) &=& -  J_{2c}(q^2_{\rm max})/2 \;,  \quad & &  
J_{1s}(q^2_{\rm max})&-& J_{2s}(q^2_{\rm max})/3   = J_{1c}(q^2_{\rm max})- J_{2c}(q^2_{\rm max})/3 \;,
  \nonumber \\
 & J_3(q^2_{\rm max}) &=& - J_4(q^2_{\rm max}) \;, \quad  & &    J_{2c} (q^2_{\rm max})& =& J_3 (q^2_{\rm max}) \, , \quad   J_{5,6s, 6c,7,8,9} (q^2_{\rm max}) =  0     \;,
\end{alignat}
which is the same number as for vector operators only. 
Above $\epq = (m_B-m_{K^*})^2$ denotes the kinematic endpoint in the $q^2$-variable. 
Many of them follow from $H_\perp(q^2_{\rm max}) = H_\perp^{\cal T}(q^2_{\rm max}) = H_\perp^{{\cal T}_t}(q^2_{\rm max}) = 0$ 
which in turn  is a consequence of parity covariance alone. We stress that the relations 
\eqref{eq:smaxlimitJi} are \emph{independent} of the dynamics.
For the CP-conjugated mode 
$ B \to  K^*(\to  K \pi)  \ell^+ \ell^-$ the $J_i$ transform as $\bar J_{1,2,3,4,7} = J_{1,2,3,4,7}$ and   
$\bar J_{5,6,8,9} = - J_{5,6,8,9}$ with all weak phases conjugated in addition. This is due to the convention that  the angle $\theta_\ell$ is defined with respect 
to the same negatively charged lepton in both $B$ and $\bar B$ decays which implies the transformations 
$\theta_l \to \theta_l - \pi$ and $\phi \to -\phi$ upon CP-conjugation.

\subsection{Kinematic endpoint:  observables in a general dimension six operator basis}

The relations \eqref{eq:smaxlimitJi} imply that many observables considered 
obey exact relations at the kinematic endpoint. 
They give rise 
to isotropic uniangular distributions\footnote{The angles are defined at the endpoint
through a limiting procedure only.}
\begin{alignat}{3}
\label{eq:l-angle}
& \frac{d^2 \Gamma}{d \cos \theta_\ell d q^2} \Big  / \left( \frac{d \Gamma}{d q^2} \right)&\; =\;& \vv \vv_\ell\left( \left(J_{1s} + \frac{J_{1c}}{2}\right) + \left(J_{6s} + \frac{J_{6c}}{2}\right) \cos \theta_\ell  + 
\left(J_{2s} + \frac{J_{2c}}{2}\right) \cos  2 \theta_\ell \right) \Big / \left( \frac{d \Gamma}{d q^2} \right) \;  &\; \to\;&  \frac{1}{2}    \;, \nonumber  \\[0.1cm]
& \frac{ d^2 \Gamma}{ d \cos \theta_K d q^2}  \Big  / \left( \frac{d \Gamma}{d q^2} \right) &\; = \;&  \vv \vv_\ell  \frac{3}{2} \left(  \left(  J_{1s} - \frac{J_{2s}}{3}\right) \sin^2 \theta_K + 
\left( J_{1c} - \frac{J_{2c}}{3}\right) \cos^2 \theta_K \right)   \Big  / \left( \frac{d \Gamma}{d q^2} \right)\;  &\; \to\;&   \frac{1}{2} \;, 
\end{alignat}
where $d \Gamma/ d q^2 = 3 \vv \vv_\ell (J_{1s} - 1/3 J_{2s})$. The interpretation of \eqref{eq:l-angle}  is that at the endpoint the $\ell \ell$- and $K\pi$-pair are in the $l=0$ (S-wave) spherically symmetric partial wave configuration.
Eq.\eqref{eq:l-angle} is also consistent with:
\begin{alignat}{3}
\label{eq:smaxlimitFL}
& F_L(q^2_{\rm max}) &\;=\;& \vv \vv_\ell \left(J_{1c} - \frac{1}{3} J_{2c}\right)  / \left( \frac{d \Gamma}{d q^2} \right)&\;=\;& \frac{1}{3}  \;, \nonumber \\[0.1cm]
 & A_{\rm FB}(q^2_{\rm max}) &\;=\;& \vv \vv_\ell \left( J_{6s} + \frac{J_{6c}}{2}\right)  / \left( \frac{d \Gamma}{d q^2} \right) &\;=\;&0 \;.
\end{alignat}
The uniangular distribution in the angle $\phi$ 
is given by
\begin{equation} \label{eq:dGdphi}
\frac{d^2 \Gamma}{d \phi d q^2} / \left( \frac{d \Gamma}{d q^2} \right)  = \frac{1}{ 2 \pi} \left( 1 +  r_\phi \cos 2 \phi  \right)  \;, \quad r_\phi \equiv \frac{ -8 J_{2s}}{9 (J_{1s} - 1/3 J_{2s})} \;,
\end{equation}
which is not isotropic in general. This is to be expected since $\phi$ is the angle 
between the two decay planes which has no special r\^ole at the kinematic endpoint. 
The result for $r_\phi$ in the general dimension six operator basis is given in appendix \ref{app:rJ}.
In the SM + SM' operator basis  
($O_{V(A)}$ and primed \eqref{eq:operators} only) one obtains in this basis  $r_\phi = -1/3 + {\cal O}(m_{\ell}^2/m_b^2)$.
This can be tested experimentally and deviations thereof could be explained by the tensor operator contributions as given in appendix \ref{app:rJ}.

The vanishing of  $J_{5,6s, 6c,7,8,9} (q^2_{\rm max})$ implies, 
besides  $A_{\rm FB} (q^2_{\rm max} )=0$, for the
CP-asymmetries $A_{i}^{(D)} \propto J_i - \bar J_i $ \cite{Bobeth:2008ij} and symmetries $S_i \propto J_i + \bar J_i $ \cite{Altmannshofer:2008dz}),
\begin{align} \label{eq:smaxlimitobs}
 A^{(D)}_{5,6,7,8,9} (q^2_{\rm max}) =0, \qquad S_{5,6,7,8,9} (q^2_{\rm max}) = 0 \, , 
\end{align}
and for the related ones $P_k^\prime$ \cite{DescotesGenon:2012zf} 
\begin{equation}
\label{eq:Pprime}
P^\prime_{5,6,8} (q^2_{\rm max}) = 0 \, , \qquad P^\prime_4(q^2_{\rm max})=\sqrt{2}\, \;.
\end{equation}
For the  transverse asymmetry $A_T^{(2)} =
%(|H_\perp|^2-|H_\parallel|^2)/(|H_\perp|^2+|H_\parallel|^2)$  
J_3/(2 J_{2s})$ \cite{Kruger:2005ep}
 $A_T^{(2)}(q^2_{\rm max}) =-1$ holds. 
The low recoil observables $H_T^{(i)}$  \cite{Bobeth:2010wg} 
\begin{alignat}{6}
  & H_T^{(1)}  
  &\equiv&  \frac{\sqrt{2} J_4}{\sqrt{-J_{2c} \left(2 J_{2s} - J_3\right)}}\;, \quad  
   & & H_T^{(1b)} &\equiv& \frac{J_{2c} J_{6s} }{2 J_4 J_5} \;, \quad   
 & &  H_T^{(2)} 
  & \equiv &  \frac{\beta_\ell J_5}{\sqrt{-2 J_{2c} \left(2 J_{2s} + J_3\right)}} \;, \nonumber \\[0.1cm]
  & H_T^{(3)} 
  &\equiv&  \frac{\beta_\ell J_{6s}}{2 \sqrt{(2 J_{2s})^2 - J_3^2}} \;,   \quad & &  H_T^{(4)} & \equiv&   \frac{2 J_8}
{\sqrt{-2 J_{2c} \left(2 J_{2s} + J_3\right)}} \,, \quad & & 
 H_T^{(5)}& \equiv&  - \frac{2 J_9}
{2 \sqrt{(2 J_{2s})^2 - J_3^2}}  \;,
 \end{alignat}
with $\beta_\ell$ defined in Eq.~(\ref{eq:norm}), obey
\begin{align} \label{eq:smaxlimitHT}
| H_T^{(1)}(q^2_{\rm max})|=1 \;, \quad H_T^{(1b)}(q^2_{\rm max}) =1\;,    \quad  \frac{H_T^{(2)}(q^2_{\rm max})}{H_T^{(3)}(q^2_{\rm max})}=1 \;,    \quad  \frac{H_T^{(4)}(q^2_{\rm max})}{H_T^{(5)}(q^2_{\rm max})}=1    \; .
\end{align}
The sign of $H_T^{(1)}$ depends on the sign of the $0$-helicity amplitude.
Note that in the second, third and fourth relation both
the nominators and the denominators are linear in $\vv$ (c.f section \ref{sec:velocity}).

Such relations have various applications. 
For instance they impose constraints on parameterizations 
of form factors and fits to decay distributions, see {\it e.g.} recently
\cite{Hambrock:2013zya} or they serve as cross checks for experimental analyses.
 We checked  explicitly that the relations Eqs.~(\ref{eq:smaxlimitJi})-(\ref{eq:smaxlimitHT}) are obeyed by the low recoil OPE results  in the most general dimension six operator basis \cite{Bobeth:2012vn}.

Note that an experimental extraction of the angular coefficients in (\ref{eq:anganal}) and derived observables requires to some extent non-local (in $q^2$) information, such as from binning or  fit shapes. While the exact size of the corresponding corrections will depend on the observable at hand, see Ref.~\cite{Hambrock:2012dg} for such a study 
and section \ref{sec:impact} for comments,
one expects them to be controlled for a sufficiently small bin at  endpoint. 
Near the endpoint the variation in $q^2$, due to possible resonances (see  section \ref{sec:expansion} for related discussions), is relatively mild.
We conclude that endpoint relations such as (\ref{eq:l-angle})-(\ref{eq:dGdphi}) are asymptotically observable.

\subsection{Small momentum  expansion (low recoil): the SM + SM' basis }
\label{sec:velocity}

We consider the observables of the previous section in the vicinity of the endpoint
within the SM + SM' basis. In addition we employ  the approximation $m_{\ell} =0$. 
This means that we only consider (axial-)vector operators $O_{V(A)}$ and the primed ones, which result
in $H^{x}_{0,\parallel,\perp}$ ($ x = L,R$) amplitudes. 
Using the results in table \ref{tab:selection} we parameterise the transversity basis near the endpoint as follows:
\begin{equation} \label{eq:kappa-expand}
H^{x}_\parallel =  - \sqrt{2}H^{x}_0 = a_{\parallel }^x + {\cal O}(\vv^2) \;, \quad H^{L,R}_\perp = a_{\perp }^x \vv  + {\cal O}(\vv^3) \;,  \quad x = L,R \; ,
\end{equation}
where $\vv$  is the absolute value of the three momentum of the $K^*$-meson in the $B$ restframe.  The relation between the three momentum $\vv$ and $q^2$ is given in \eqref{eq:speedD}. More details as well as
phenomenological parameterisations for low 
and high $q^2$ including further terms in the $\vv$-expansion are briefly sketched in appendix \ref{app:q2expansion}. 

In this work we neglect  CP-violation, because in $B \to K^*$ transitions direct CP-violation is small in the SM and there is no experimental evidence for it presently. We remark that
CP-violating effects to the parameterisations could be included by simply introducing the 
analogous parameters 
for CP-conjugated quantities, allowing to form CP-averaged observables and CP-asymmetries.
Expanding up to linear order in $\vv$ one obtains: 
\begin{alignat}{12}
\label{eq:Jv}
& J_{1s} &\;=\;& \frac{3}{4} \AAA \;, \quad & & J_{1c} &\;=\;& \frac{1}{2} \AAA \;,  \quad &  & J_{2s} &\;=\;& \frac{1}{4} \AAA \;, \quad  &  & J_{2c} &\;=\;& - \frac{1}{2} \AAA \;, \quad  &  & J_{3} &\;=\;& -\frac{1}{2} \AAA \;, \quad  &  & J_{4} &\;=\;& \frac{1}{2}  \AAA \;, \nonumber \\
 &  J_{5} &\;=\;&  \RRR \vv \;, \quad  &  & J_{6c} &\;=\;& 0   \;, \quad  &  & J_{6s} &\;=\;& 2 R \vv   \;, \quad & & J_{7} &\;=\;&  0 \;, \quad  &  & J_{8} &\;=\;& \frac{1}{2} \III \vv \;, \quad  &  & J_{9} &\;=\;& - \III \vv \;,
\end{alignat}
with further corrections of \emph{relative} order ${\cal O}(\vv^2)$. Here we use
the following:
\begin{alignat}{2}
\label{eq:ARI}
& \AAA &\equiv\;&  |a_{\parallel}^L|^2 + |a_{\parallel}^R|^2 =  \frac{1}{2} (|a_{\parallel}^V|^2 + |a_{\parallel}^A|^2)  \;, \nonumber \\[0.1cm]
&  \RRR&  \equiv\;&
 \text{Re}[a_\parallel^{L} a^{L*}_{ \perp} - a_\parallel^{R} a^{R *}_{ \perp}]  
 =  -\frac{1}{2}  \text{Re}[a_\parallel^{V} a^{A*}_{ \perp} + a_\parallel^{A} a^{V *}_{ \perp}]     \;,  \nonumber \\[0.1cm]
&  \III&  \equiv\;&  \text{Im}[a_\parallel^{L} a^{L*}_{ \perp} + a_\parallel^{R} a^{R *}_{ \perp}] =   \frac{1}{2} \text{Im}[a_\parallel^{V} a^{V*}_{ \perp} + a_\parallel^{A} a^{A *}_{ \perp}]   \;,
\end{alignat}
and $a^{V/A}=a^R \pm a^L$.
The relations \eqref{eq:Jv} are compatible with \eqref{eq:smaxlimitJi} in the limit $\vv \to 0$ and match the expressions in chapter III.B of Ref.~\cite{Bobeth:2012vn} to the given order. 
We stress that \eqref{eq:Jv} is though independent of any dynamical assumptions.
Observables  linear in $\vv$ are:

\begin{alignat}{4}
& A_{\rm FB} &\;=\;& \frac{J_{6s}-J_{6c}/2}{3 J_{1s} - J_{2s}}   =   \hat \RRR \vv \;, \quad 
& & P_5' &\;=\;& 
 \frac{J_5}{2 \sqrt{ -J_{2c} J_{2s} }} =  \,     \sqrt{2}  \hat \RRR \vv \;, \nonumber  \\[0.2cm]
&  \quad S_8 &\;=\;&  \frac{4/3 J_8}{3 J_{1s} - J_{2s}}  =    \, \frac{1}{3 }  \hat \III \vv \;, \quad & &  S_9 &\;=\;& \frac{4/3 J_9}{3 J_{1s} - J_{2s}}  =  
-   \,  \frac{2}{3}  \hat \III \vv \, ,
\end{alignat}
where
\begin{equation}
\hat \RRR \equiv \frac{\RRR}{\AAA}\;, \qquad  \hat \III \equiv \frac{\III}{\AAA} \;.
\end{equation}
Consequently, in the vicinity of the endpoint the following relations hold,
\begin{align}
\frac{ P_5' }{ A_{\rm FB}} =\sqrt{2} + {\cal O}(\vv^2) \; ,\quad \qquad \frac{ S_8 }{S_9} = - \frac{1}{2} + {\cal O}(\vv^2) \; , \quad \qquad \frac{H_T^{(4)}}{H_T^{(2)}} = \frac{\hat \III}{\hat \RRR} + {\cal O}(\vv^2) \, .
\end{align}

We compare the endpoint predictions with existing data from BaBar~\cite{BaBarLakeLouise}, CDF~\cite{HidekiICHEP2012}, LHCb~\cite{Aaij:2013iag,Aaij:2013aln,Aaij:2013qta}, 
ATLAS~\cite{ATLAS:2013ola} and  CMS~\cite{CMS:cwa} in table \ref{tab:vergleich-all}. 
\begin{table}[h]
\begin{center}
\begin{tabular}{ l || c c c c  c c||  c c c  c }
 & $F_L$  & $S_3$  & $^a$$P_4'$ & $S_7$ & $^b$$ P_5'/A_{\rm FB}$ & $^b$$S_8/S_9$ &  $^a$$A_{\rm FB}$ & $P_5'$ & $^a$$S_8$ & $S_9$ \\ \hline
endpoint & $1/3$ & -$1/3$ & $ \sqrt{2}$ & $ 0 $  & $\sqrt{2}$ & $-1/2$& $ \hat \RRR \vv$  & $\sqrt{2} \hat \RRR \vv$ & $1/3 \hat \III \vv $ & $ -2/3\hat \III \vv$  \\ \hline
$B \to K^* $ & $0.38 \pm 0.04$ & -$0.22 \pm 0.09$ & $0.70^{+0.44}_{-0.52}   $ &$0.15^{+0.16}_{-0.15}$ & $1.63 \pm 0.57$ & $-0.5 \pm 2.2$ &$  -0.36 \pm 0.04 $ & $-0.60^{+0.21}_{-0.18}$ & $-0.03 \pm 0.12$ & $0.06^{+0.11}_{-0.10}$  \\
$B_s \to \phi $ & $0.16^{+0.18}_{-0.12}$ & $0.19^{+0.30}_{-0.31} $ & -- & --  & -- &-- & --  & -- & -- & --
\end{tabular}
\end{center}  
\caption{\small Comparison of endpoint predictions (second row) for angular observables (first row) to the current world average  in the available endpoint-bin $q^2 \in [16,19]\GeV^2$ (LHC-experiments) or otherwise $q^2 \in [16 \GeV^2, q^2_{\rm max}]$
for $B \to K^* \ell^+  \ell^-$ decays (third row; our error weighted average, systematic and statistical uncertainties are added in quadrature). The last row gives corresponding data for $B_s \to \phi \mu^+ \mu^-$ decays \cite{Aaij:2013aln} (LHCb only).
$^a$ Experimental values adopted to theory definitions as in  \cite{Bobeth:2012vn} ($A_{\rm FB}$), \cite{Altmannshofer:2009ma} ($S_i$)
and \cite{DescotesGenon:2012zf} ($P^\prime_k$).
$^b$ with symmetrized errors and assuming gaussian error propagation. Note, $S_3=1/2 (1-F_L) A_T^{(2)}$.  All corrections are 
of relative order ${\cal O}(\vv^2)$ (for $m_{\ell} =0$) by virtue of parity covariance.}
\label{tab:vergleich-all}
\end{table}
The data are consistent with the endpoint relations.
The largest deviation is at the 2$\sigma$-level  in the statistics-limited $B_s \to \phi \mu^+ \mu^-$ analysis  \cite{Aaij:2013aln}.  The prediction for the ratio
$ P_5'/A_{\rm FB}$ is satisfied at 1$\sigma$. Since both measurements of $S_8$ and $S_9$ are currently consistent with zero, the corresponding 
ratio comes with a large uncertainty. The perfect agreement of the present experimental central value with the endpoint prediction has therefore to be seen as accidental.

Investigating the observables linear in $\vv$ we find from table \ref{tab:vergleich-all} 
\begin{align} \label{eq:IR-fit}
\hat R=( -0.67 \pm 0.07) \GeV^{-1}, \quad \hat I=( -0.17 \pm 0.27) \GeV^{-1}, \quad
\hat I/\hat R=0.25 \pm 0.40\, .
\end{align}
The bin-averaged 3-momenta are  $\vv_{\rm bin} =0.55 \GeV$ for $q^2 \in [16,19]\GeV^2$ 
and $\vv_{\rm bin} =0.52 \GeV$ for $q^2 \in [16 \GeV^2, q^2_{\rm max}]$.
The measured value of $\hat R$ is in accordance with its SM prediction, $\hat R_{\rm SM}=( -0.73^{+ 0.12}_{-0.13}) \GeV^{-1}$\cite{Bobeth:2012vn}. 
In the limit where the strong and the weak phases do not depend on the polarisation, or both phases are negligible, $\hat I=0$ holds. 
In the SM  $\hat I =0 $ (which is consistent with \cite{Bobeth:2010wg}) since there are no sizeable weak phase differences and as argued in the next section the leading strong phases are polarisation independent indeed.  The current experimental results in \eqref{eq:IR-fit} 
are therefore  consistent with the SM.

\subsection{The validity of small momentum expansion and $c \bar
c$-resonances \label{sec:expansion}}

To evaluate the performance of the $\vv$-expansion we discuss in section \ref{sec:impact}
the (non)-impact of charmonium resonances in the low recoil region, and compare suitable observables at low recoil to data (section \ref{sec:kappaatwork}). In section \ref{sec:non-fac}
we discuss possible experimental probes of non-factorizable contributions in this region.

\subsubsection{The (non-)impact of the $cc$-resonances at low recoil}
\label{sec:impact}

Charm-resonances  \cite{PDG} contribute to  $B \to K^{(*)} \ell \ell$ via 
$B \to K^{(*)}  (c \bar c \to \ga^* \to  \ell \ell)$ and are visible in the local spectrum 
starting from $q^2  = m_{J/\Psi}^2 \simeq 9.6 \GeV^2$ onwards.
While such effects have been known to the theory community {\it e.g.,} \cite{Ligeti:1995yz,Kruger:1996cv,Ali:1999mm},
the pronounced structure observed recently  in the low recoil region in
the $B \to K \mu^+ \mu^-$ dimuon spectrum \cite{Aaij:2013pta} reinforces one
to think about the most suitable binnings.
Here  we would like to point out that for observables composed out of
certain ratios of 
HAs the situation is simpler since factorizable (universal)
contributions, which are supposedly leading, drop out as
they are polarisation independent. 
To show this we write the HAs, neglecting right handed currents, as follows:
\begin{eqnarray}
\label{eq:schema}
H_i^V &=& F^V_i(q^2)\Big( 1+ L^{\rm fac,c}(q^2) + L_i^{\rm
n\text{-}fac,c}(q^2) + ... \Big)  \;, \nonumber \\
H_i^A &=& F^A_i(q^2)\Big(1 + ...\Big)  \;, \qquad\qquad\qquad\qquad\qquad     \quad   \; i=\perp,\parallel ,0 \;, 
\end{eqnarray} 
where $F_i^{V(A)}(q^2) \propto C_{V(A)} f_i(q^2)$, $L^{\rm fac,c}$ and
$L_i^{\rm n\text{-}fac,c}$ denote the non-resonant ({\it e.g.}~$F_\parallel^{V(A)} \propto  C_{9(10)} A_1(q^2)$ in the conventions of \cite{Bobeth:2012vn}),  factorizable charm and
non-factorizable charm contribution, respectively. The ellipses stand for all other terms.
The  charm loop with no gluon reconnecting to any other part of the diagram
is part of $L^{\rm fac,c}$  (naive factorization) and the charm loop with gluons
emitted into the vector meson final state constitutes a part of $L_\la^{\rm n\text{-}fac,c}$. 
Similar to the cancellation of (universal) short-distance coefficients, universal charm-contributions
 $L^{\rm
fac,c}$ drop out in observables which are ratios of certain combinations of HAs composed of the sum
of squares  as $(H^L_i H^{L*}_j + H^R_i H^{R*}_j)/(H^L_l H^{L*}_k + H^R_l
H^{R*}_k )$, where $i,j,k,l=\perp,||,0$ \cite{Bobeth:2010wg}.
This can also be seen by defining
\begin{eqnarray}
\label{eq:lucky0}
x_{ij}(a) &\equiv&  ( H^V_i H^{V*}_j + a H^A_i H^{A*}_j) \nonumber    \\[0.1cm] 
&= & (1+a)( H^L_i H^{L*}_j +  H^R_i H^{R*}_j) + (1-a) ( H^L_i H^{R*}_j +  H^R_i H^{L*}_j) \nonumber    \\[0.1cm] 
&\propto&  f_i(q^2) f_j^*(q^2) ( |C_V|^2 |1 + L^{\rm fac,c}(q^2)|^2 +  a | C_A |^2) + {\cal O}(L^{\rm n-fac,c})      \;,
\end{eqnarray} 
where $a$ is a complex number.
Thus for observables of the form,
\begin{equation}
\label{eq:lucky}
\Phi( x_{ij}(a), .. , x_{lk}(a)  ) \;, \quad     \Phi(b x, .. ,  b y ) =  \Phi(x,..,y)  \quad \text{with }b \text{ a number and }     i,j,k,l =\perp,\parallel ,0 \;,
\end{equation}
the $L^{\rm fac,c}$-contribution effectively drops out.
Examples of observables of the form \eqref{eq:lucky}  are:
\begin{align} 
\label{eq:FL}
F_L&  \equiv
 \frac{|H_0^L|^2 + |H_0^R|^2}{\sum_{X=L,R} ( |H_0^X|^2+
|H_\perp^X|^2+|H_\parallel^X|^2 )}\, , \\  \label{eq:AT2}
    A_T^{(2)} & \equiv
  \frac{|H_\perp^L|^2 + |H_\perp^R|^2-(|H_\parallel^L|^2+|H_\parallel^R|^2)}
       {|H_\perp^L|^2 + |H_\perp^R|^2+|H_\parallel^L|^2+|H_\parallel^R|^2}\; , \\
       P_4^{\prime} & \equiv  \frac{ \sqrt{2} {\rm Re} (H_0^L H_\parallel^{L*}+H_0^R H_\parallel^{R*})}{\sqrt{(|H_\perp^L|^2 + |H_\perp^R|^2+|H_\parallel^L|^2+|H_\parallel^R|^2)(|H_0^L|^2 + |H_0^R|^2)}} \label{eq:P4prime}\; ,
\end{align}
where the latter requires for $L^{\rm fac}$ to cancel that there are no sizeable weak phase differences between
the $H_0$ and $H_\parallel$ HAs. Corresponding CP-asymmetries are given in \cite{Bobeth:2011gi,Bobeth:2012vn}. We recall that
the presence of  right-handed currents ($C_+ \neq C_-$  in table \ref{tab:selection})  would affect the cancellations 
in observables like $A_T^{(2)}$. Whereas this poses no problem for the SM where $C_+ \simeq C_-$   it would make precise interpretations of beyond the SM right-handed currents in the low recoil region more difficult.

While the evaluation of non-factorizable contributions $L_i^{\rm
n\text{-}fac,c}$,
{\it e.g.}~\cite{Ball:2006eu}, remains an important task,
we stress that the endpoint relations obtained in this work imply that
these non-factorizable contributions become polarisation independent at the endpoints, see
\eqref{eq:smaxlimitcalD}, {\it i.e.,}
\begin{equation}
L_\la^{\rm n\text{-}fac,c}(\epq) =   L^{\rm n\text{-}fac,c}   \;,\quad  \la = 0,\pm1,\parallel,\perp\;.
\end{equation}

\subsubsection{The small momentum expansion at work for $F_L$ and $A_T^{(2)}$}
\label{sec:kappaatwork}

The endpoint relations are subject to  corrections away from the endpoint which we
have parameterised in Eq.~\eqref{eq:kappa-expand} in terms of a $\vv$-expansion.
The uncertainty estimate ${\cal O}(\vv^2)$ is in general hampered by the 
$c \bar c$-resonance effects.  However, as pointed out in the previous section
 such effects essentially drop out in observables which are of the form \eqref{eq:lucky}.
Hence the corrections to the HAs 
can be taken to be of the order  ${\cal O}(\vv^2/q^2)$  (where ${\cal
O}(q^2) = {\cal O}(m_B^2)$ and $m_{K^*}$ cannot enter the denominator
because the HAs have got to
be smooth in the limit $m_{K^*} \to 0$ \cite{Dimou:2012un}).

Neglecting the small $q^2$-dependence of $C_9^{\rm eff}$ from quark loops as well as terms ${\cal O}(C_7/C_9^{\rm eff})$, 
$H_{\parallel}(q^2) \propto A_1(q^2)$ and we can estimate the $q^2$-correction
at the beginning of the last bin by looking at 
$R_{\rm bin} = A_1(16 \GeV^2)/A_1(\epq)$.   
One obtains $R_{\rm bin} \simeq 0.85$  for both the extrapolated LCSR results in \cite{Ball:2004rg} as well as the recent lattice results \cite{Horgan:2013hoa}. Thus $15\%$ (which has to be compared with the naive parametric value $\vv^2/q^2 = 0.06$) can be taken as a measure of the error at the end of the endpoint bin. The averaged error over the bin should be 
only half the size thus one might expect the parameterisation to hold within $10\%$.
 
To further substantiate this  we study $F_L$ \eqref{eq:FL} and $A_T^{(2)}$ \eqref{eq:AT2}, which 
belong to type \eqref{eq:lucky}.  $F_L$  interpolates between
$F_L(0) = 0$ where the longitudinal mode decouples completely
and $F_L(\epq) = 1/3$ \eqref{eq:smaxlimitFL} where all polarisations are equally probable.
In the intermediated regime it assumes a maximum around $F_L = {\cal O}(0.8-0.9)$ as a result of  the zero of the $A_{\rm FB}$ 
and the equivalence theorem c.f. appendix \ref{app:FLasymptotics}.   $A_T^{(2)}$
interpolates between $A_T^{(2)}(0) \simeq 0$ due to $H^{V,A}_{+} \simeq 0$  by virtue of the $V$-$A$-interactions and $A_T^{(2)}(\epq) = -1$.

The  data on $F_L$ and $A_T^{(2)}$ for $B \to K^* \ell
\ell$  is compiled in  table \ref{tab:FL}. The result   
$F_L|_{\rm LHCb}=0.523 \pm 0.005 \pm 0.010$ for $B_s \to J/\Psi \phi$ \cite{Aaij:2013oba} shows as expected a similar pattern to $B \to J/\Psi K^*$ decays.
The change of $F_L$ observed in the next-to endpoint
bin $[14.18,16] \GeV^2$ is within the expected order ten percent range of $F_L=1/3$ 
and the sign of the change is consistent with the previously mentioned theoretical
considerations. The changes in $A_T^{(2)}$ versus $q^2$ are  larger and come presently with a large experimental uncertainty.  A significant change is, however, expected as the observable is forced, as we argued, to change rather abruptly from $A_T^{(2)}(\epq ) =  -1$ to $A_T^{(2)}|_{\text{low } q^2} \simeq 0$.

\begin{table}[h]
\begin{center}
\begin{tabular}{c||c|c|c|c|c||c}
& $J/\Psi$   &   & $\Psi(2S)$ & & & $\chi_{c1}$ \\
$q^2$ [GeV$^2$] & 9.59 & $[10.09,12.86]$  & 13.59 &  $[14.18,16]$ &
$[16,X]^\dagger$ & 12.33\\
\hline
$F_L$ & $0.566 \pm 0.007$ & $0.48 \pm 0.05$  & $0.48 \pm 0.05$ &
$0.38 \pm 0.05$ & $0.38 \pm 0.04$& $0.77 \pm 0.08$ \\
$A_T^{(2)} $$^*$ & $-0.008 \pm 0.025$ & $-0.36  \pm 0.30$  & $0.15 \pm 0.17$ &
$0.07 \pm 0.25$ & $-0.64 \pm 0.27$& $-0.74 \pm 0.35$ \\
\end{tabular}
\end{center}  
\caption{\small $F_L$ and $A_T^{(2)}$ measured in $B \to K^* \ell^+ \ell^-$ decays on
\cite{Aubert:2007hz,Aaij:2013cma} and off-resonance
\cite{BaBarLakeLouise,HidekiICHEP2012,Aaij:2013iag,ATLAS:2013ola,CMS:cwa}
(our error weighted average, systematic and statistical uncertainties are
added in quadrature). $^\dagger$$X=19$ for LHC experiments and $X=q^2_{\rm
max}$ otherwise. The last column corresponds to $B \to K^*
(\chi_{c1} \to J / \Psi \gamma)$ decays \cite{Aubert:2007hz} for which
large non-factorizable corrections are expected. $^*$Our evaluation of $J/\Psi, \Psi(2S)$ and  $\chi_{c1}$  using Eq.~(\ref{eq:AT2}).}
\label{tab:FL}
\end{table}

\subsubsection{Low recoil OPE versus non-factorizable corrections \label{sec:non-fac}}

\begin{figure}[ht]
\begin{center}
\includegraphics[width=0.4\textwidth]{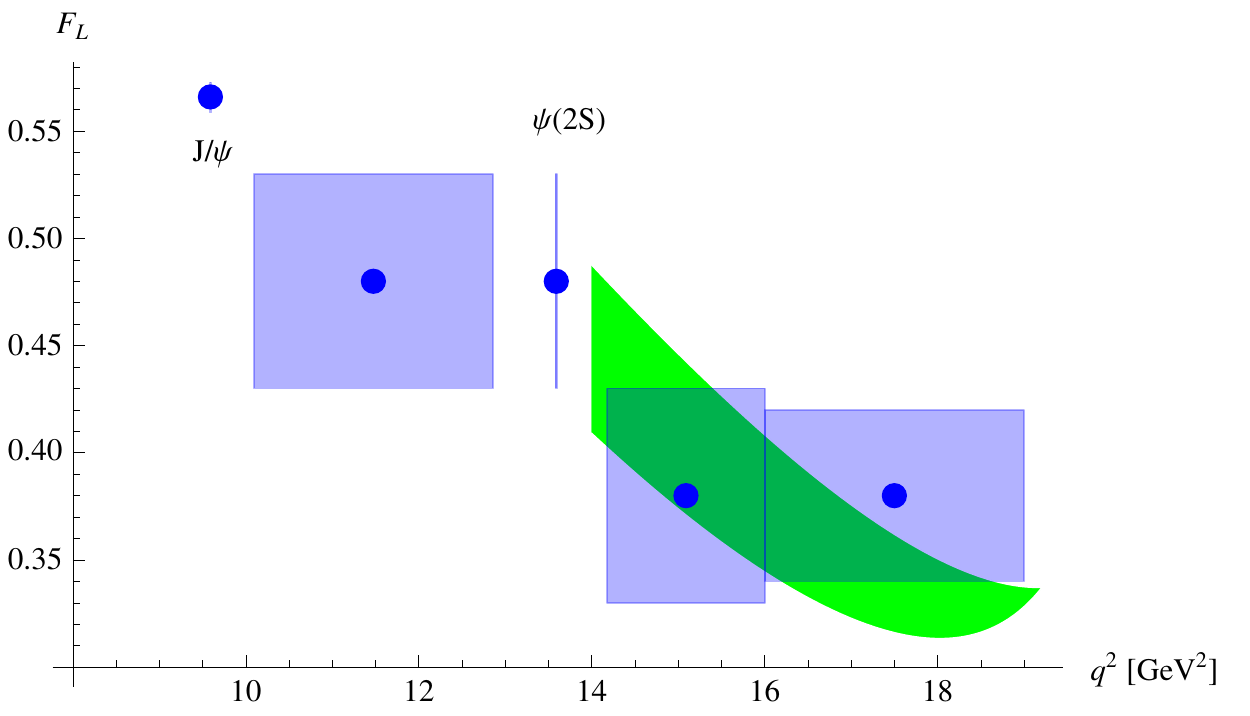}
\includegraphics[width=0.4\textwidth]{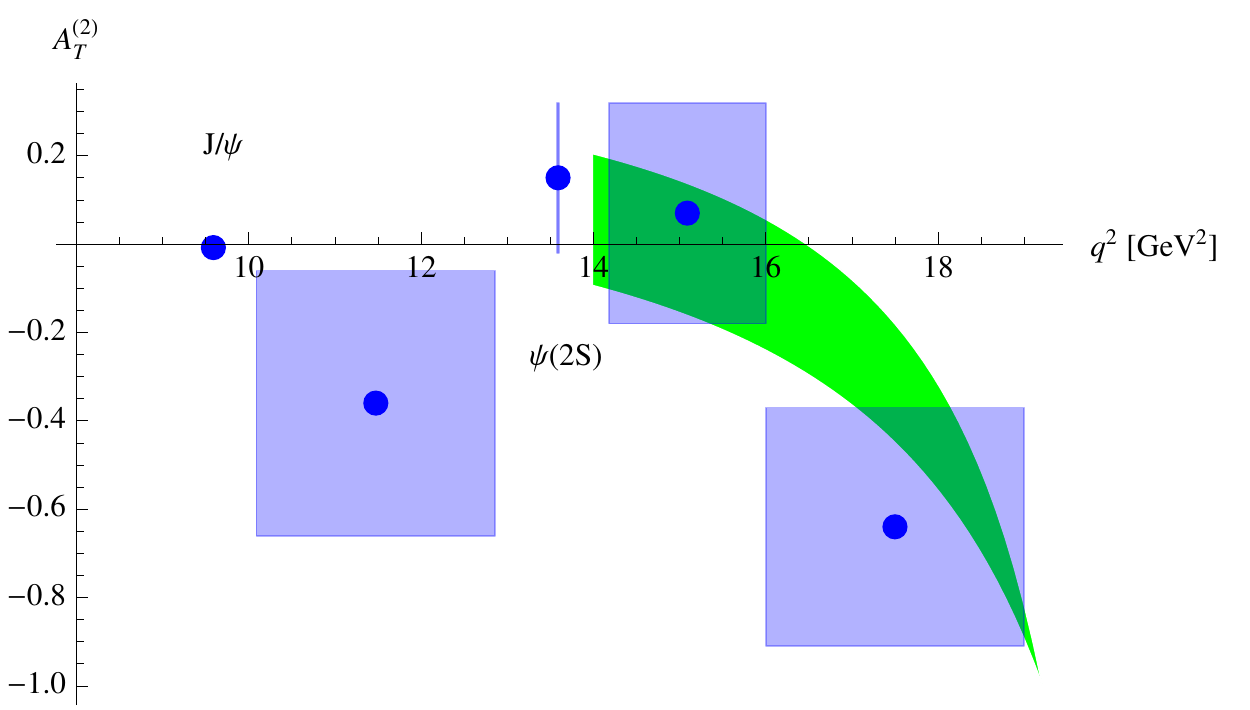}
\end{center}
\caption{\small $F_L$ (left-hand side) and $A_T^{(2)}$ (right-hand side) in the intermediate and  low recoil  region in $B
\to K^* \ell^+ \ell^-$ decays on and off $c\bar c$-resonance, see Table
\ref{tab:FL}. The (green) band corresponds to  a model-independent
prediction based on the leading OPE within the SM basis from
Ref.~\cite{Hambrock:2013zya}, see text.}
\label{fig:FL}
\end{figure}

The absence of estimates of the non-factorizable contributions $L_i^{\rm
n\text{-}fac,c}(q^2)$ in \eqref{eq:schema} are a limiting factor for precision predictions
in the low recoil region.  Turning this argument around the sensitivity to such effects can be used to obtain experimental  information on the size of the $L_i^{\rm n\text{-}fac,c}(q^2)$.
In this context  $F_L$ and $A_T^{(2)}$ are useful
as  factorizable $c\bar c$-effects effectively drop out. 
In Fig.~\ref{fig:FL} we show $F_L$ and $A_T^{(2)}$ from $B \to K^* \ell^+ \ell^-$ decay
data on (blue points) and off $c\bar c$-resonance (blue boxes) as in table \ref{tab:FL}.
For the predictions in the SM basis we use the OPE at leading order in $1/m_b$ for which $F_L$ and $A_T^{(2)}$ are    form factor based \cite{Bobeth:2010wg}.
Under these assumptions ratios of form factors can be obtained from the data
and used to predict $F_L$  and $A_T^{(2)}$. 
The (green) band corresponds
 to such a recent determination \cite{Hambrock:2013zya}.
The plots are made with the 'SE2 LEL'  fit at $1 \sigma$; other
parameterisations give comparable results at low recoil, see
\cite{Hambrock:2013zya} for details.

For $F_L$ and $A_T^{(2)}$ the  OPE predictions agree well 
at current precision. 
 This supports that non-factorizable, non-universal contributions
are subdominant at  low recoil. (For estimations of OPE uncertainties, see \cite{Grinstein:2004vb,Beylich:2011aq}.)
As argued in the previous section the increase of $F_L$ away from the endpoint  region is in the ballpark 
of what can be expected by the $\vv$-expansion. Moreover the same is true for extrapolated  large recoil factorization results
\cite{Bobeth:2010wg}, suggesting that non-factorizable (non-universal)
contributions are subdominant in ratios at larger recoil as well.

Two further remarks on resonances are in order.  
First the $\chi_{c1}$ as well as $\chi_{c0}$ and $\chi_{c2}$ are known
to receive  large non-factorizable corrections and one  therefore cannot
expect extrapolations
from the endpoint
to be trustworthy. 
This is indeed reflected in a rather high value of $F_L(\chi_{c1})$, see
table \ref{tab:FL}. Second, in general observables for resonances cannot be expected to be close 
to their neighbouring off-resonance counterparts since they couple to a different set of
operators. 
The fact that the resonant $c \bar c$-contributions for $F_L$
are almost at par with the non-resonant findings, see Fig.~\ref{fig:FL}, is therefore non-trivial from the viewpoint of dynamics and supports the kinematic interpretation in terms of the equivalence theorem.
The exception being the endpoint region  where the predictions become increasingly exact.
For $F_L$ and $A_T^{(2)}$ this also persists
to some degree for lower $q^2$, as can be inferred from  Fig.~\ref{fig:FL}, since by the equivalence theorem and the $V$-$A$ interactions both observables are expected to follow  a certain pattern in that region. 
Further compilation and discussion of $F_L$ in the context of the non-leptonic decays 
to two vector mesons is discussed in section \ref{sec:SVV}.

We end with a discussion on $P_4'$ which figures as well as $F_L$ and $A_T^{(2)}$ amongst the observables of type \eqref{eq:lucky} which are not affected by $L^{\rm fac,c}$ in \eqref{eq:schema}.
The value of $P_4^\prime=-0.18^{+0.54}_{-0.70}$  measured in the next-to endpoint  bin $[14.18,16] \, \mbox{GeV}^2$ \cite{Aaij:2013qta} deviates by
$3 \sigma$ and is larger than the expected ${\cal{O}}(\vv^2/q^2)$ away 
from its endpoint value $\sqrt{2}$ \eqref{eq:Pprime}.  This appears to be conflicting
the conclusions drawn from $F_L$ and $A_T^{(2)}$ regarding the $\vv$-expansion and/or the size of the non-factorizable charm contributions although the uncertainties in $P_4'$ data are currently sizeable. Comparing Eqs.~(\ref{eq:FL}), (\ref{eq:AT2}) to (\ref{eq:P4prime}), 
it is tempting to introduce order one phases between $H_0$ and $H_\parallel$ while
keeping ratios of  moduli of the HAs approximately unchanged.
This choice, however, would affect the ratio $ P_5'/A_{\rm FB}$, schematically $\propto {\rm Re}(H_0 H^*_\perp)/{\rm Re}(H_\perp H^*_\parallel)$. Its value in the bin $[14.18,16] \, \GeV^2$, $ P_5'/A_{\rm FB} =1.66 \pm 0.55 $ \cite{Hambrock:2013zya}, is in agreement with the endpoint predictions  and the value in the endpoint bin, see table \ref{tab:vergleich-all}.
Further measurements of 
$B \to K^* \ell \ell$ angular observables using different binnings as well
as on-resonance
measurements of asymmetries can help to clarify these issues and to
quantify $ L_\la^{\rm n\text{-}fac,c}(q^2)$.

\subsection{Applications to other vector modes}
\label{sec:other}

The endpoint relations obtained for the $B \to K^* (\to K \pi) \ell^+ \ell^-$-template apply to other modes of the same Lorentz structures and spin configurations. 
 Examples of rare $S \to V (\to P_1 P_2) \ell^+ \ell^-$ decays are 
\begin{eqnarray} \nonumber
B \to \rho (\to \pi \pi) \ell^+ \ell^- , \quad B_s \to \phi (\to  KK) \ell^+ \ell^- , \quad B_s \to K^*(\to K \pi) \ell^+ \ell^- , \\ \nonumber
B_c \to D_s^*( \to D_s  \pi^0) \ell^+ \ell^- , \quad B_c \to D^* (\to  D \pi) \ell^+ \ell^- , \\
D \to \rho (\to \pi \pi) \ell^+ \ell^-, \quad D_s \to K^*(\to K \pi)\ell^+ \ell^- \,, \label{eq:fcncmodes}
\end{eqnarray}
as well as lepton flavor violating ones $S \to V \ell^+ \ell^{\prime -}$, where $\ell \neq \ell^\prime$.
The predictions equally apply to the charged current decays  $S \to V (\to P_1 P_2)  \ell \nu$:
\begin{eqnarray}
\label{eq:ccmodes}
 \nonumber 
 B \to D^*(\to D \pi) \ell \nu , \quad B_s \to D_s^*( \to D_s  \pi^0) \ell \nu , \quad B_s \to K^*(\to K \pi) \ell \nu , \quad B \to \rho (\to \pi \pi) \ell \nu \,, \\ \nonumber
  B_c \to \psi(3770) (\to  D D) \ell \nu , \quad B_c \to D^*(\to D \pi) \ell \nu ,  \\
 D \to \rho (\to \pi \pi) \ell \nu, \quad D_{(s)} \to K^* (\to K \pi) \ell \nu \, , \quad
D_s \to \phi (\to K K) \ell \nu \, .
\end{eqnarray}
The distributions for the $\ell \nu$ final states differ from 
the $\ell \ell$ final state by terms of order ${\cal O}(m_{\ell}/m_{c,b})$. 
The predictions equally apply to higher states of the above $V$'s with the same flavor content if kinematically allowed.

\section{$B \to X_j \ell \ell$-type decays}
\label{sec:otherthanV}

In section \ref{sec:zero} and section \ref{sec:higher} we discuss meson decays to spin zero and
$\geq 2$, respectively.

\subsection{Spin 0 \label{sec:zero}}

The decay  $B \to K \ell \ell$ 
is of the simplest kind since there is no polarisation tensor for the spin zero $K$.  The scalar ($O_{S(P)}$ and primed) amplitude $H$ can be non-vanishing as there are 
simply no Lorentz indices that need to be saturated. The vector ($O_{V(A)}$ and primed) amplitudes allow for $\gpol^*(\lambda_B) \cdot  \pI$, and hence 
$H_\la \propto {\cal O}(\vv)$ for $\la = 0,\pm 1$ and $H_t \propto {\cal O}(\vv^0)$. The latter is 
accompanied by a factor of $m_{\ell}$ when acting on the lepton-bilinear and one might therefore write 
$H_t^{\rm eff} \propto {\cal O}(m_{\ell} \vv^0)$.
 The tensor amplitudes ($O_{\cal T}$ and primed) allow for  the
invariant $\gpol^*_{\mu\nu}(\la_{B_1} ,\la_{B_2}) q_{r}^{\mu} q_{r'}^\nu $, where
$\lambda_{B_1}= \bar \lambda_{B_2}$, by virtue of \eqref{eq:rule}, which results in  
 $H_{\la \la'} = 0$ and $H_{t\la} \propto {\cal O}(\vv)$. In summary we obtain:
\begin{equation}
\label{eq:Hscalar}
H \propto {\cal O}(\vv^0) \;, \quad  H_\la \propto {\cal O}(\vv) 
\;, \quad  H_t^{\rm eff} \propto {\cal O}(m_{\ell} \vv^0) \;, \quad H_{\la \la'} = 0  \;, \quad H_{t\la} \propto {\cal O}( \vv)
  \;,
\end{equation}
where here $\la,\la' = 0,\pm 1$. The power-counting is in agreement with explicit calculations at large and low recoil,  \cite{Bobeth:2007dw,Bobeth:2012vn}, respectively.  

The decay rate $\bar B \to \bar K \ell^+ \ell^-$,   see Eq.~(\ref{eq:norm}) and footnote \ref{foot:root} for normalization, for a general dimension six operator basis can be written as \cite{Bobeth:2007dw}
\begin{align}
\label{eq:K}
\frac{1}{\vv \beta_\ell} \frac{ d^2 \Gamma}{d \cos \theta_\ell d q^2} \;=\;  a_\ell+ b_\ell \cos \theta_\ell + c_\ell \cos \theta_\ell^2        \;,
\end{align}
where the lepton angle $\theta_\ell$ is defined as in $\bar B \to \bar K^* \ell^+ \ell^-$ decays. The relations \eqref{eq:Hscalar} imply  $b_\ell,c_\ell(\epq)=0$ at the endpoint (here $\epq =(m_B-m_K)^2$) and  imply isotropicity:
\begin{alignat}{3}
\label{eq:K-angle}
& \frac{d^2 \Gamma}{d \cos \theta_\ell d q^2}   / \left( \frac{d \Gamma}{d q^2} \right)&\;=\;& \vv \beta_\ell \left(  a_\ell+ b_\ell \cos \theta_\ell + c_\ell \cos \theta_\ell^2  \right)  / \left( \frac{d \Gamma}{d q^2} \right) \;  &\; \to \;&  \frac{1}{2}    \;, 
\end{alignat}
where $d \Gamma/ d q^2 =  \vv \beta_\ell 2( a_\ell+ c_\ell/3) $.
Eq.\eqref{eq:K-angle} is also consistent with:
\begin{alignat}{3}
\label{eq:smaxlimitFH}
& F_H(q^2_{\rm max}) &\;=\;& \vv \beta_\ell 2  \left(a_\ell+c_\ell \right)  / \left( \frac{d \Gamma}{d q^2} \right)&\;=\;& 1 \;, \nonumber \\[0.1cm]
 & A_{\rm FB}(q^2_{\rm max}) &\; = \;& \vv \beta_\ell  b_\ell  / \left( \frac{d \Gamma}{d q^2} \right) &\;=\;&0 \;.
\end{alignat}
$F_H$ is called flat term in the distribution. We note that \eqref{eq:Hscalar} implies 
 $a_\ell(q^2_{\rm max})|_{\rm SM}=0$. Non-vanishing endpoint contributions arise from 
(pseudo-) scalar operators only.

The endpoint relations for HAs and angular distribution for $B \to K^*_0(1430)  \ell \ell$ decays
hold analogously with $C_+ \to C_-$ since $J^P(K^*_0) = 0^+$. The decay with 
$K^*_0(1430) \to K \pi$ has been considered as a background  process for $B \to K^* (\to K \pi) \ell\ell$ decays with different angular decomposition \cite{Becirevic:2012dp}.  Fortunately, for the low recoil region this 
effect is small since the
kinematic endpoint $q_{\rm max}^2 = (m_B - m_{K^*_0(1430)})^2 \simeq (3.85 \GeV)^2$ 
barely overlaps with the signal region above $m_{\Psi(2S)}^2$.

Comparing the $\kappa$-counting in Eq.~(\ref{eq:Hscalar}) for $B \to K \ell \ell$ with 
the one for $B \to K^* \ell \ell$ in table \ref{tab:selection}, one observes that they are different,
{\it e.g.} $H_\la^{B \to K} \propto {\cal O}(\vv)$ whereas
$H_\la^{B \to K^*} \propto {\cal O}(\vv^0)$. On the other hand, there is a relation between 
$B \to K \ell\ell$ and decays to a longitudinally $K^*$, $B \to K^*_L \ell \ell$ at low $q^2$
to leading twist and assuming that one of the chiralities dominates, such as in the SM (c.f. section 5.1.3. \cite{Lyon:2013gba}). The absence of a relation at high $q^2$, where the twist expansion does not make sense a priori anyway, can also be understood from the fact that at the endpoint
there is a democracy amongst the helicity directions and since the $K$-meson has not got any analogue of the $\pm1$ helicity direction the relation has got to break down.

\subsection{Higher spin}
\label{sec:higher}

We consider decays $S \to X_j \ell^+ \ell^-$ (and $S \to X_j \ell \nu$), where $X_j$ denotes a hadron of spin $J$.  For a  $S \to X_{J}  X'_{J}$-type decay 
the longitudinal {\it i.e.} $0$-helicity polarisation fraction assumes the value 
$F_L = 1/(2J+1)$ \cite{RZ13}, provided the HAs do not vanish by a 
selection rule.
The situation for  $S \to X_j \ell^+ \ell^-$ in the effective Hamiltonian approach 
\eqref{eq:operators} is entirely different. 

These statements are illustrated through the  $B \to K^*_2(1430)  \ell^+ \ell^-$-decay. 
We  follow the same type of analyses as in section \ref{sec:tensor},
restricting ourselves to the SM + SM' basis.
\emph{Through invariants}: we construct an invariant out of
a spin $2$ and $1$ polarisation tensor $\gpol_{\mu\nu}^*(\bar \la_C)$ and $\gpol^*_\mu(\la_B)$ 
as well as $(\pI)_\mu \propto \gpol_{\mu}^*(t)$. The only possible candidate is 
$(P_1)_{\la_B \la_C} =   \gpol_{\mu\nu}^*(\bar \la_C) \gpol^{*\mu}(\la_B) \gpol^{*\nu}(t)$,
which vanishes at the endpoint because of transversity $\pI \cdot \gpol(\la_R,\pI) = 0$. 
More precisely one finds that:
\begin{equation}
\label{eq:spin2}
H_\la  \propto (P_1)_{\la \la} \propto {\cal O}(\vv)  \; , \qquad H_t \propto (P_1)_{t0}\propto  {\cal O}(\vv^2) \;,
\end{equation}
where $\la =0, \pm$ and we have  used  that $0 = \la_A = \la_B + \bar \la_C$.
\emph{Through CGC:} The following product is of relevance: 
$( \R{2}_B \times \R{1}_C)  \times \R{1}_{\pI} = 1 \cdot  \R{1} + ..$ which implies that
 $H_\la \propto   C^{121}_{0 \la \bar \la}$.
Either method yields the following ratios between HAs:
\begin{equation}
\label{eq:H2}
 H_{\bar 2 } \;:\;   H_{\bar 1 } \;:\;  H_{0 } \;:\;   H_{1 } \;:\;  H_{2 } \;= \; 0  \;:\;  1  \;:\;  \frac{-2}{\sqrt{3}}  \;:\;  1  \;:\;  0 \;.
\end{equation}
It is clear that the corresponding uniangular 
angular distribution is \emph{not}  isotropic at the endpoint. 
From \eqref{eq:H2} one obtains 
$F_L(\epq) = 2/5$ and not $1/5$ which would be consistent with isotropicity. The fact that $H_\la \sim {\cal O}(\vv)$ and  $H_t \sim {\cal O}(\vv^2)$ \eqref{eq:spin2} can be seen 
from \cite{Li:2010ra} at the level of form factor contributions. 
Taking the endpoint limit in \cite{Li:2010ra} we find agreement with \eqref{eq:H2}
up to sign differences which could be due to different conventions of polarisation vectors.

Generalisations of this result are evident. For example for integer  $J \geq 2$ 
and with vector operators $O_{V(A)}$-transition operators,
the $H_\la \propto {\cal O}(\vv^{J-1})$ (with $H_\la =0$ for $|\la| > 1$) and $H_t \propto {\cal O}(\vv^J)$. The former can also be understood in another way. 
Since the $(X_j \ell\ell)$-state is at least in a $l=J\!-\!1$-wave (with $l$ being the total orbital angular momentum) the result $H_\la \propto \vv^{J-1}$ is therefore expected. 
The generalisation to higher dimensional operators with higher derivatives is not
straightforward because of the timelike polarisation.

\section{Endpoint relations and non-leptonic decays}
\label{sec:range}

In this section we extend the discussion of the introduction  to what extent  endpoint relations apply  to non-leptonic decays. 
A few sample decay modes are listed at the end of this section.
 
A hybrid case between the non-leptonic and the semileptonic case arises when 
 the lepton pair is produced 
via a hadron, {\it e.g.,}~$B_s \to \phi(1680)  \Psi(2S)( \to \ell \ell)$ for which we expect endpoint symmetries to work.
Note that the masses are tuned [$(m_{B_s},m_{\phi(1680)},m_{\Psi(2S)})  = ( 5.36,1.68,3.68) \GeV$] such that the leptons are likely to be coming from the hadronic state. The decay $B \to K^* \Psi(2S)$, which is in the wider vicinity of the endpoint c.f. Fig.~\ref{fig:FL}, gives some support of this statement since $F_L$ is reasonably close to its endpoint value of 
$1/3$.
The essence 
of these examples are that the production mechanism of the lepton pair is immaterial and it is the quantum numbers of the actual final states which determine the helicity structure at and near the endpoint.

What happens in a decay which is purely non-leptonic such as $B \to K^*(\to K \pi) p \bar p$ for example is  more complicated as in general one expects virulent final state interactions.
Consider that the decay $B \to K^* p \bar p$ proceeds as depicted in Fig.~\ref{fig:ABC} (with $A=B$, $C=K^*$ 
and $B_{1,2} = p,\bar p$), then there is a configuration where $\vv= |\vec{q}_{K^*}| = 0$ 
which implies $\vec{q}_{p} = -\vec{q}_{\bar p}$ for which we can expect isotropicity 
(in $\theta_p$). 
The crux is though that the $K^*$ then decays into $K \pi$ and that these particles 
will interact strongly with the $p \bar p$-pair. 
In the extreme case of inelasticity this leads to a different final state (say $p K \to \Sigma \pi$) or the $K$ and the $p$ might simply exchange momentum, both of which changes the kinematics in general. Thus we are led to conclude that endpoint symmetry holds,
if one restricts oneself to the subset of decay configurations from $B \to K^*(\to K \pi) p \bar p$  decays for which  $\vec{q}_{p} = -\vec{q}_{\bar p}$. The latter could be realized if the proton pair does not interact significantly with the $K^*$-meson.
In essence we expect endpoint symmetries to play a r\^ole on a subset 
of configurations which is defined by $\vec{q}_C =0$ and $q_C^2 = m_C^2$.
 
Decays that could be studied experimentally consist of  $(B_1 B_2)$ pairs which either scalars such as $(\pi\pi), (\pi K)$ or 
long-lived spin-1/2 baryons such as $(p \bar p)$ or $( \Lambda\bar \Lambda)$. 
For the baryons several modes have been observed to date,
$B \to D^{0*} p \bar p$ \cite{Aubert:2009qz},
$B \to K^* \Lambda \bar \Lambda$ \cite{Chang:2008yw} and
$B \to K^* p \bar p$. For the latter  there are already $K^*$-polarisation  measurements, which however do not cover the endpoint region \cite{Chen:2008jy}.
The decay $B_{(s)} \to J/\Psi p \bar p$ has been searched for at LHCb \cite{Aaij:2013yba}.
Corresponding branching ratios are ${\cal{B}}(B \to K^* \Lambda \bar \Lambda, K^* p \bar p)=\mbox{few}\times {\cal{O}}(10^{-6})$ and ${\cal{B}}(B \to D^{0*} p \bar p)={\cal{O}}(10^{-4})$ \cite{PDG}.

\subsection{Hadronic $S \to V_1 V_1$ and  qualitative remarks on $F_L$}
\label{sec:SVV}

Decays of the type $S \to V_1 V_2$ are generically not in an endpoint configuration. Here we take
the quantity $u \equiv ((m_{V_1} + m_{V_2}) / m_S)^2$  as 
a measure of the distance from the endpoint. Towards the endpoint $F_L|_{u \to 1} \to 1/3$ \eqref{eq:smaxlimitFL} (with available phase space vanishing), and naively we   expect $F_L|_{u \to 0} \to 1$ by virtue of  
the equivalence theorem as explained in appendix \ref{app:FLasymptotics}. The difference 
to the discussion in the appendix is though that neither of the vector mesons is observed through $V \to \ga^* \to \ell \ell$ and the cancellations leading to the zero of the $A_{\rm FB}$ are therefore absent.  In table \ref{tab:FLfun} we collect data on selected $B \to V_1 V_2$ decays which support
the anticipated pattern in $F_L(u)$.

\begin{table}[h]
\begin{center}
\begin{tabular}{l || l l l l l }
$B \to V_1 V_2$ & $B^0 \to D_s^{*+} D^{*-}$ & $B^0 \to J/\Psi K^{*0}$  & $B^0 \to D^{*+} D^{*-}$ & 
$B^0 \to  D_s^{*+} \rho^-$     & $B^0 \to \rho^{+} \rho^{-}$ \\ \hline
$u$ & $0.61$ & $0.57$ & $0.58$ & $0.28$ &  $0.09$ \\
$F_L$ \cite{PDG} & $0.52 \pm 0.06$ & $0.570 \pm 0.008$ & $0.624 \pm 0.031$ & $0.84 \pm 0.03$ & $0.977 \pm 0.026$ 
\end{tabular}
\end{center}  
\caption{\small Selected data on $B \to V_1 V_2$ decays that illustrate $F_L|_{u=1} = 1/3$ and  $F_L|_{u \to 0} \to 1$. Whereas the former is exact the latter limit is based on qualitative arguments which can be invalidated by the specific dynamics, see text.}
\label{tab:FLfun}
\end{table}

\begin{figure}[ht]
\begin{center}
\includegraphics[width=0.4\textwidth]{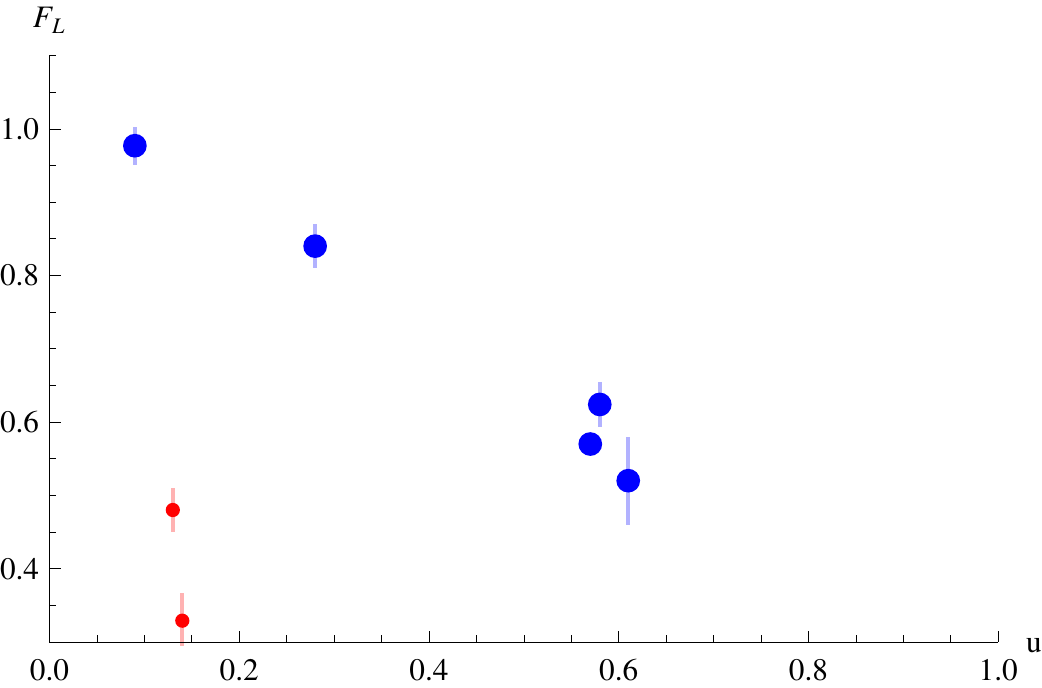}
\end{center}
\caption{\small $F_L$ for $B \to V_1V_2$ examples (blue points) for  
$F_L|_{u=1} = 1/3$ and  $F_L|_{u \to 0} \to 1$,  as in table \ref{tab:FLfun}.  The smaller (red) points correspond to decays with different weak dynamics, see text. }
\label{fig:FLfun}
\end{figure}

Specific dynamics including $V$+$A$ admixture can, however, invalidate  
the kinematic picture. 
For example  $F_L(B^0 \to \phi K^{*0})_{u = 0.13} = 0.480 \pm 0.030$ \cite{PDG} differs considerably from  $F_L(B^0 \to \rho^+ \rho^-)_{u = 0.09} = 0.977 \pm 0.026$ despite their close
$u$-values. 
An explanation was pointed out in  \cite{Kagan:2004uw}\footnote{For a more complete classification we refer the reader to reference \cite{Beneke:2006hg}.}  where it  was observed that in  penguin dominated decays  formally subleading 
 contributions in $1/m_b$, known as weak annihilation, are numerically large and can in principle accommodate the experimental results.  Other examples of the same kind, recently measured by the LHCb collaboration, are
 $F_L(B_s \to \phi \phi)_{u = 0.14} = 0.329 \pm 0.033 \pm 0.017$  \cite{Aaij:2013qha} and 
 $F_L(B_s \to \phi K^*)_{u = 0.13} = 0.51 \pm 0.15 \pm 0.07$  \cite{Aaij:2013gga}.
 In Fig.~\ref{fig:FLfun} we show the decays from  table \ref{tab:FLfun} supporting the
 kinematic $F_L$-pattern (blue points) as well as the rather precisely measured penguin modes where it does not work (smaller red points).
 
 The uncertainties of the weak annihilation theory predictions  are  rather large 
 in QCD factorization because of  endpoint (infrared) divergences for which 
 a cut-off has to be introduced. 
This precludes  conclusions on possible new physics signals  in those observables.
Further analysis is beyond the scope of this paper.

Predictions for $D \to V_1 V_2$ decays  are more stable as the  
corresponding $u$-values are larger and in some cases very close to  
the endpoint kinematics, see Table \ref{tab:u-DVV}.
While current data on $F_L$ in these modes except for $F_L(D^0 \to \rho^0 \rho^0) =0.690 \pm 0.074$ \cite{PDG} are not precise enough for  
a comparison with the endpoint prediction and asymptotics,
we  encourage further study.

\begin{table}[h]
\begin{center}
\begin{tabular}{l || c c c c c }
$D \to V_1 V_2$  & $ \rho \rho$ & $K^* K^* $ & $K^* \rho$ &
$\phi \rho $     & $\phi K^*$ \\ \hline
$u(D_0,D_\pm)$ & $0.68$ & $0.92$ & $0.80$ & $0.92$ &  --\\
$u(D_s)$ & $0.61$ & $0.83$ & $0.72$ & $0.83$ &  $0.95$
\end{tabular}
\end{center}
\caption{\small $u$-values for $D \to V_1 V_2$ decays. The columns  
(rows) correspond to
final (initial)  states.}
\label{tab:u-DVV}
\end{table}

\section{Summary}
\label{sec:conclusions}

The key results of this work are the endpoint relations for the helicity amplitudes \eqref{eq:summary} and angular observables of a decaying spinless particle into a vector meson and two leptons, Eq.~(\ref{eq:smaxlimitJi}). 
Constraints on observables 
are summarised in Eqs.~(\ref{eq:l-angle})-(\ref{eq:smaxlimitHT}).
The physics origin of the relations is the absence of direction
at the point of zero recoil. The helicity relations apply analogously to
charged-current decays (\ref{eq:ccmodes}), including for example $B \to D^* \ell \nu$. In the case of di-neutrino modes $B \to K^* \nu \bar \nu$ \cite{Altmannshofer:2009ma} 
 and all modes with $\ell=\nu$ from Eq.~(\ref{eq:fcncmodes})
$F_L=1/3$  constrains the  shape of the decay distribution. 
As the  symmetry  is independent of the dynamics of the decay but rather depends on the spin of the external states, it  applies as well to non-leptonic decays of the type $B \to K^* p \bar p$ provided that 
the $K^*$ does not interact significantly with the proton pair. We investigated the vicinity of 
the endpoint through an expansion in the three momentum of the $V$-meson in the $B$-cms in section \ref{sec:velocity}. 

The endpoint relations are beneficial in guiding experimental searches.  
Existing  endpoint data is  found to be consistent  with the endpoint symmetries, see Table \ref{tab:vergleich-all}. The agreement 
validates that backgrounds are under control. In addition,
Eq.~(\ref{eq:smaxlimitFL})  can be used to remove backgrounds from other resonant (and non-resonant) contributions to $S \to P_1P_2 \ell^+ \ell^-$ from intermediate particles other than spin $1$.  An important example is the S-wave contributions to $B \to K \pi \ell^+ \ell^-$, recently addressed in Refs.~\cite{Becirevic:2012dp,Matias:2012qz,Blake:2012mb} for the low $q^2$ region.

New physics opportunities arise from  in  $d \Gamma/d \phi$, Eq.~(\ref{eq:dGdphi}), the distribution between the two final decay planes, which is sensitive  to tensor operator contributions,  as well as the slopes in the vicinity of the endpoint.  
The latter, denoted by $\hat \RRR$, $\hat \III$ in (\ref{eq:Jv},\ref{eq:ARI}),  are universal and appear in several observables. The overdetermination of $\hat{\RRR}$ and $\hat{\III}$ provides a model-independent check. 
Current extractions from {\it e.g.,} $A_{\rm FB}$ and $P_5^\prime$ are
consistent with each other and with the SM.
Second
order corrections in the momentum expansion are not universal but for observables like $F_L$, $A_T^{(2)}$ and $P_4'$, in which resonance effects mostly drop out (c.f. section \ref{sec:impact}), the relative change with respect to the endpoint prediction serves as an additional test. 
While data on $F_L$ and $A_T^{(2)}$ change only moderately and accordingly in 
the next to endpoint bin, $P_4'$ \cite{Aaij:2013qta} deviates substantially and 
escapes interpretations within factorization \cite{Descotes-Genon:2013wba,Altmannshofer:2013foa,Hambrock:2013zya,Beaujean:2013soa}. 
It will be interesting to see how this develops with the forthcoming analysis of LHCb's  $3 \mbox{fb}^{-1}$ data
set, and with the future data taking of the Belle II experiment.

Additionally our findings explain the observed universal helicity structure  
of $B \to K^* \ell^+ \ell^-$, and the ones in (\ref{eq:fcncmodes}),
in the SM basis \cite{Bobeth:2010wg} (c.f. appendix \ref{app:comment})
manifest in the low recoil OPE \cite{Grinstein:2004vb}.  
Furthermore we found that
$F_L$ follows a kinematic pattern imposed by the endpoint symmetry and the
equivalence theorem (c.f. appendix \ref{app:FLasymptotics}),
at the exception of channels with numerically enhanced weak annihilation
contributions, see Fig.~\ref{fig:FLfun}.
The corresponding data on non-leptonic decays is compiled and commented  on in section \ref{sec:SVV}.
We investigated several generalisations, including decays into a spin-zero and spin-2 meson plus two fermions. Further decays maybe studied at and in the vicinity of their endpoint, towards a more precise interpretation of
future weak decay data.\\

{\bf Acknowledgements:}  
We are grateful to Greig Cowan, Markus Hopfer, Sebastian J\"ager, J\"urgen K\"orner, James Lyon, Franz Muheim, Matt Needham,  Stefan Schacht and Wei Wang for useful discussions.
RZ  acknowledges the support of an advanced STFC fellowship.
This work is supported by the Deutsche Forschungsgemeinschaft (DFG) within research unit
{\bf \sc FOR} 1873.

\appendix
\numberwithin{equation}{section}

\section{Conventions for polarisation vectors}
\label{app:pol}

Our conventions of the physical polarisation tensors are those of Jacob and Wick \cite{JW59} 
and are the same as in \cite{RZ13}.
For a polarisation vector for a particle at rest we have:
\begin{eqnarray}
\label{eq:JW}
\gpol(\pm)  =  (0, \pm 1, i,0)/\sqrt{2} \;, \quad \gpol(0) = (0,0,0,1) \quad \text{and} \quad  \gpol(t) = (1,0,0,0 )  \;.
\end{eqnarray}
Throughout this work  we use $(+,-) \leftrightarrow (1,\bar 1) \leftrightarrow (1,-1) $   interchangeably for a spin 1 polarisation index.
The identification with the polarisation tensors $\alpha, \beta,\gamma$ of the particles $A$, $B$ and $C$, respectively, taken to be $J=1$, is as follows: 
\begin{eqnarray}
\gpol(\pm) &=&      \al(\pm) =   \be(\pm) =   \ga(\mp)  \;, \\[0.1cm] \nonumber
\gpol(0) &=&  \al(0) =  \be(0) =  \ga(0)  \;,
\end{eqnarray}
where the last two equalities are valid at the endpoint as can be seen from the explicit parameterisation to given below. The scalar products are 
\begin{equation}
\label{eq:SP}
\gpol(\la_1) \cdot \gpol^*(\la_2) = - \delta_{\la_1 \la_2} \;, \quad \gpol(\la_1) \cdot \gpol(\la_2) =  -  (-1)^{\la_1} \delta_{\la_1 \bar  \la_2}  \;. 
\end{equation}

The only complication is that the quantization axis of the $C$-particle is opposite to 
the  ones of the $A$- and $B$-particles. This implies a helicity flip 
as indicated in the equation above.  Furthermore in the Jacob-Wick phase convention 
the helicity dependent phase is chosen to be unity \cite{JW59,Haber:1994pe}. 
We emphasise that the conventions \eqref{eq:JW} are consistent with 
the Condon-Shortly phase convention which are standard for 
the  CGC \cite{PDG}.

The power counting of the $\vv$-corrections in the vicinity of the endpoint in the decay $A \to (B_1B_2) C$ can be verified by using the explicit parameterisation of the vectors in the $A$-cms:
\begin{alignat}{6}
\label{eq:polis}
&\pB  &\;=\;& q = ((\pB)_0,0,0, \vv),  \quad  \quad & &  \be(0)&\;=\;&(\vv,0,0,(\pB)_0)/ \sqrt{\pB^2} \;, \quad & & \be(t)&\;=\;&\pB /\sqrt{\pB^2} \;,  \nonumber \\
&\pC &\;=\;&  p =  ((\pC)_0,0,0,-\vv) \;,  \quad 
& & \ga (0)& =& (-\vv,0,0, (\pC)_0)/\sqrt{\pC^2}  \;,  \quad  & &   \ga(\pm)&\;=\;&\be(\mp) = \gpol(\mp) \;.
\end{alignat}
Above $(\pB)_0 \equiv \sqrt{\pB^2 + \vv^2}$, $(\pC)_0 \equiv \sqrt{\pC^2 + \vv^2}=m_A-(\pB)_0$ and  the physical polarisation vectors are transverse $\pB \cdot \be(\la)=0$, $\pC \cdot \ga(\la)=0$   for $\la=0, \pm$.  For assessing the $\kappa$-expansion we choose 
the $A$- and $C$-particle to be on-shell $\pA^2 = m_A^2$ and $\pC^2 = m_C^2$.
The variable $q_B^2 = \epq + {\cal O}(\vv^2)$   is  $\vv$-dependent through Eq.~(\ref{eq:speedD}). The distance to the kinematic endpoint is $\delta q^2 \equiv  (m_A-m_C)^2 - \pB^2 = {\cal O}(\vv^2)$. With these conventions the relevant scalar products read:
\begin{alignat}{2}
& \be(0) \cdot \pC &\;=\;& \vv  m_A  / \sqrt{\pB^2}={\cal{O}}(\kappa) \,,  \nonumber \\
& \be( \pm ) \cdot \pC &\;=\;& 0  \;,\nonumber \\[0.1cm]
& \be(t) \cdot \pC  &\;=\;& (m_A^2 - q_B^2 - m_C^2)/(2 \sqrt{\pB^2})  ={\cal{O}}(\kappa^0)  \;,  
\end{alignat}
and
\begin{eqnarray}
\label{eq:gade}
 \be(\la_B) \cdot  \ga( \la_C)=  \begin{cases}
1 & \la_B  = \bar  \la_C = \pm   \\[0.1cm]
(-\vv^2 - \sqrt{\vv^2 + \pB^2} \sqrt{\vv^2 + m_C^2})/( \sqrt{\pB^2} m_C )  & \la_B  =  \la_C = 0  \\[0.1cm]
0 & \text{otherwise}\end{cases} \;.  \nonumber  \\
\end{eqnarray}

\section{The low recoil OPE and non-factorizable corrections  in the  
light
of endpoint relations}
  \label{app:comment}

   It was observed that in the SM+SM' basis
  the effective Wilson coefficients (EWC) $\CPCPV^{\rm eff}(q^2) \equiv
C^{\rm eff}(q^2) \pm C^{\rm eff '}(q^2)$ are independent of the $V$
mesons' polarisation \cite{Bobeth:2010wg,Bobeth:2011nj,Bobeth:2012vn}.
In the transversity basis this amounts to
  \begin{align} \nonumber
B \to V \ell \ell:& \quad   \HA_{0,\parallel} = \CPV^{\rm eff}(q^2)
f_{0,\parallel}(q^2) \;, \quad \HA_{\perp} = \CPC^{\rm eff}(q^2) f_
\perp(q^2) \;,  \\
B \to P \ell \ell:  & \quad H  = \CPC^{\rm eff}(q^2) f(q^2)  \;,
\label{eq:ansatz1}
\end{align}
where $f_i$ ($i=0, \perp, \parallel$) are the usual
polarisation-dependent $B \to V$ form factors and
  $f$ denotes the corresponding $B \to P$ form factor.
Generally the expression \eqref{eq:ansatz1} is subject to non-
factorizable
corrections
\begin{equation}
  f_\la(q^2) \to   f_\la(q^2) (1+ \epsilon_\la(q^2) ) \;, \quad
\epsilon_\la(q^2) =  {\cal{O}}(\alpha_s/m_b,[ {\cal{C}}_7/{\cal{C}}_9]/
m_b)  \;,  \quad  \la = 0,\pm 1 \; .
\end{equation}
We would like to point out that the endpoint relation imply degeneracy  
of
the corrections   at the endpoint
\begin{align} \label{eq:smaxlimitcalD}
\epsilon_\la(q^2_{\rm max})  \equiv  \epsilon \;,      \quad  \la =
0,\pm 1,\parallel,\perp   \,,
\end{align}
with \eqref{eq:summary} already enforced by the form factors $f_\la$,
  $f_\perp(q^2_{\rm max})=0$ and $f_\parallel (q^2_{\rm max})=
\sqrt{2} f_0(q^2_{\rm max})$ as used {\it e.g.,}  in
Ref.~\cite{Hambrock:2013zya}.
Below we give two alternative viewpoints
on why the EWC are polarisation independent in the low recoil region.
\begin{itemize}
\item
   For the true Wilson coefficients, which in particular are $q^2$-
independent universality follows from factorization, a property of the  
OPE.
   The points to be
discussed are contributions from off-shell photons and
   quark loops {\it e.g.}, $\bar b s \bar c c$-operators which lead
to so-called  charm loops.
For the former the improved Isgur-Wise relations
\cite{Isgur:1990kf,Grinstein:2002cz} are instrumental,
which render the relative size of the electromagnetic dipole to
(axial-)vector operator contributions
universal. The origin of this feature are the QCD equations of
motions, further discussed in
\cite{Hambrock:2013zya}.
The reason why \eqref{eq:ansatz1} holds for the charm loops as well is
that those
loops factorize and do  therefore not introduce a new dependence on
the polarisation per se.
More precisely the charm loop corresponds to the vacuum polarisation  
which
is  proportional
to  $g_{\al \be}$ and $q_\al q_\be$ structures where $q$ is the
external  momentum
going through the loop. The latter structure vanishes when it is
propagated by the photon
to a vector current by the conservation of the latter. The  $g_{\al
\be}$-structure does
not change the polarisation.
Thus the factoriztion into a form factor and direct evaluation of a $
\bar \ell \gamma_\mu \ell$ matrix element implies \eqref{eq:ansatz1}.
\item  Another viewpoint is that Eq.~(\ref{eq:summary})  enforces within
\eqref{eq:ansatz1}
  relations between the  form factors.  To make this explicit
  consider a generic ansatz as
\begin{equation}
\HA_\la = C_\la (q^2) f_\la(q^2)   + C'_{\la} (q^2) f_{\bar \la}(q^2)  
\;,
\quad  \la =
0,\pm 1    \;.
\end{equation}
By virtue of parity covariance (c.f. table \ref{tab:selection}) the form
factors obey $f_{\la}' = - f_{\bar \la}$.
Since the form factors obey the same endpoint relations as the  
amplitudes
\eqref{eq:summary}
$f_0(\epq) = - f_1(\epq) = - f_{\bar 1}(\epq)$ it follows that the
combination
$C_\la(\epq) -C'_\la(\epq)  = C$  is polarisation independent.  
Moreover,
since the relation must hold irrespective of the ratio of $V$-$A$ and
$V$+$A$ operators this implies complete degeneracy at the endpoint.
One might wonder whether the equality of the EWC holds at low recoil as
well as at the endpoint. The answer to that is affirmative as the EWC  
does
not know about the location of the endpoint.  We could for instance  
apply
the formalism to $B \to K^*(1410) \ell \ell$
in which case the endpoint would lower from  $(m_B-m_{K^*})$ to
$(m_B - m_{K^*(1410)})$
with the same EWC. Thus one gets
\begin{equation}
\label{eq:0pm}
\HA_{\la} = C(q^2) f_\la(q^2) -C'(q^2) f_{\bar \la} (q^2) \;,
\end{equation}
which is equivalent to the expression in \eqref{eq:ansatz1} after
identifying $C^{(')}(q^2)$ with $C^{{\rm eff}(')}(q^2)$.
\end{itemize}

  At last we would like to point out that the results (\ref{eq:ansatz1})
are valid independent of the
chirality of the lepton interaction vertex.
One could restore $\HA_{i} \to \HA_{i}^{L,R}$  with $L,R$ referring to
$\ell \to \ell_{L,R}$.
The fact that in general the Wilson coefficient of the transverse HA
is different from the other can also be inferred from the fact that
the corresponding
form factor vanishes at the endpoint $f_\perp(\epq) =0$ and evades a
constraint.
For models with $|C| \gg |C'|$, such as the SM,
degeneracy in all transversity directions $0,\parallel,\perp$ is
effectively
attained.

\section{Phenomenological parameterisations for helicity amplitudes}
\label{app:q2expansion}

Within the  SM + SM' basis (neglecting $H_t$ which is suppressed by $m_\ell$) 
one may parameterise $H_\la$, $\la = 0,\parallel,\perp$ to be used in fits to experimental data. We suggest here a phenomenological, local parameterisation adapted to low and high $q^2$ without working out all details\footnote{\label{footnote:redundant} One needs to take into account that the angular coefficients  $J_i$ are bilinears of the HAs. This leads to a
$U(2)$-symmetry and reduces the number of free parameters by four
\cite{Egede:2010zc}.}.

The $B \to V \ell \ell$ decay rate scales like $d \Gamma \propto  \vv \vv_\ell d q^2 d(\text{angles})$ where $\vv$ and $\vv_\ell$ denotes the absolute value of the three momentum of the $K^*$ and $\ell$-particles, respectively. The former is given in Eq.~\eqref{eq:speedD} and the latter is given by 
 \begin{equation} \label{eq:norm}
 \vv_\ell \equiv \sqrt{\frac{\la(q^2,m_\ell^2,m_\ell^2) }{4 q^2}} = \frac{1}{2} \sqrt{q^2} \beta_\ell \; , \quad
 \beta_\ell \equiv \sqrt{1-4 \frac{m_\ell^2}{q^2}} \; .
  \end{equation}
In counting and parameterising higher orders in $\vv^2$ or $q^2$ it is therefore important 
to keep track of both $\vv$ and $\vv_\ell$. In the conventions of reference \cite{Bobeth:2012vn} this  is taken care of by a global factor 
$ \sqrt{\vv \vv_\ell}$ in the HAs. Below we do not write the $ \sqrt{\vv \vv_\ell}$-prefactor explicitly.

\subsection{Low recoil -- high $q^2$} 
 
We extend the $\kappa$-expansion  \eqref{eq:kappa-expand} to:
\begin{alignat}{2} 
\label{eq:kappa-expand2}
-\sqrt{2} & H^{x}_0 &\;=\;&  \sqrt{ \epq/q^2} (a_{0}^{x} + b_{0}^{x} \vv^2 +  \frac{c_0}{q^2} +.. ) (1+ \sum_r \Delta_0^{(r)}(q^2))   \;,  \quad   x = L,R  \;, \nonumber \\[0.1cm]
& H^{x}_\perp &\;=\;& \vv ( a_{\perp }^x  + b_{\perp }^x \vv^2 + \frac{c_\perp}{q^2} +   .. )(1 +\sum_r \Delta_\perp^{(r)}(q^2))   \;,  \nonumber  \\[0.1cm]
& H^{x}_\parallel &\;=\;&  (a_{\parallel }^x + b_{\parallel }^x \vv^2 + \frac{c_\parallel}{q^2} + .. )(1 +\sum_r \Delta_\parallel^{(r)}(q^2))    \;,
\end{alignat}
where the coefficients $a,b$ and $c$ are complex (see, however,  footnote \ref{footnote:redundant}), the function $\Delta^{(r)}(q^2)$ take into account effects 
of resonances and the ellipses stand for higher orders in $\vv$. 
Locally the $\Delta$-function  can be approximated by a Breit-Wigner ansatz
$\Delta^{(r)}_i(q^2)  = \delta_i(q^2) /(q^2 - m_r^2 + i\Gamma_r m_r)$ where   $m_r$, $\Gamma_r$ are the corresponding masses and decay widths. In the factorization approximation 
$\delta_i(q^2) = \delta_i$ is a constant and deviations thereof are a measure of non-factorizable contributions.

The endpoint constraint \eqref{eq:summary} implies:
\begin{equation}
 (a^x_0+c_0/\epq)  (1 + \sum_r  \Delta_0^{(r)}(q^2_{\rm max})  ) 
 = (a^x_\parallel + c_\parallel/\epq) (1 + \sum_r  \Delta_\parallel^{(r)}(q^2_{\rm max})  )   \;, \quad x = L,R\;.
\end{equation}
Since the  $a,c,\Delta$-terms stem in general from different operators in the effective Hamiltonian, barring fine-tuning,  the
endpoint relation must be satisfied for each term separately, that is
\begin{equation}
a^x_0= a^x_\parallel \; , \quad c_0= c_\parallel  \; , \quad \Delta_0^{(r)}(q^2_{\rm max}) \, = \Delta_\parallel^{(r)}(q^2_{\rm max}) \; ,
\end{equation}
consistent with Eq.~(\ref{eq:smaxlimitcalD}). 

A few remarks are in order. First of all, the parameterisation \eqref{eq:kappa-expand2}
aims at an efficient  phenomenological description and is supposed to hold locally, unlike the OPE.
The factor $1/\sqrt{q^2}$ in front of $H_0$ originates from the normalisation of the 
polarisation vector \eqref{eq:polis}. The  resonance terms $c$ as well as $\Delta^{(r)}$ originate from 
photons and thus couple vector-like, $c^R = c^L$, $\Delta^R=\Delta^L$ and we 
have therefore suppressed in (\ref{eq:kappa-expand2}) their chirality labels. Instead of $\vv^2$ one could also expand in 
$\delta q^2 \equiv (\epq-q^2)$ c.f. \eqref{eq:k2q2}.
 The deviation of ratios from one, $[\Delta_\la^{(r)}(q^2) ]/[\Delta_{\la'}^{(r)}(q^2) \,  ]\neq 1$ is a $q^2$-dependent measure of non-factorizable 
 contributions to  observables which depend on 
ratios of HAs. 

The remaining HAs that arise in a general dimension six effective Hamiltonian can be parameterised in an analogous  but simpler manner as $c\bar c$-resonances  are absent:
\begin{alignat}{3} 
\label{eq:kappa-expandBSM}
& H_X  &\;=\;&  \vv/\sqrt{q^2} (a_{X} + b_{X} \vv^2 +.. ) 
 \;,   \quad  \, & &  \mbox{for} ~H_X=H_\perp^{\cal{T}},H_\perp^{{\cal{T}}_t},H_t \;, \nonumber \\[0.1cm]
 & H_X &\;=\;&  \sqrt{ \epq/q^2} (a_{X} + b_{X} \vv^2 +.. ) 
 \;,  \quad    \, & &  \mbox{for} ~H_X=H_\parallel^{\cal{T}},H_\parallel^{{\cal{T}}_t} \;, \nonumber \\[0.1cm]
  & H_X &\;=\;&   a_{X} + b_{X} \vv^2 +..  
 \;,  \quad    \, & &  \mbox{for} ~H_X=H_0^{\cal{T}},H_0^{{\cal{T}}_t}  \;, \nonumber \\[0.1cm]
 & H&\;=\;&   \vv /\sqrt{ \epq} (a_{} + b_{} \vv^2 +.. )  \;. & & 
\end{alignat}
Note that the pseudo-scalar contribution is absorbed as usual in the vector one by the equations of motion.

\subsection{Large recoil -- low $q^2$}

At large recoil  a similar parameterisation to the one at low
recoil given in the previous subsection can be employed\footnote{
It was suggested to apply $q^2$-expansion in the large recoil region
\cite{Reece,Petridis}.}:
\begin{alignat}{2}
\label{eq:kappa-expand2loq2}
-\sqrt{2}&H^{x}_0 &\;=\;&  \sqrt{ \epq/q^2} (\al_{0}^{x} + \be_{0}^{x}
q^2 + .. )  (1+ \sum_r \Omega_0^{(r)}(q^2))  \;,  \quad   x = L,R \;,
\nonumber \\[0.1cm]
& H^{x}_\perp &\;=\;& \vv( \al_{\perp }^x  + \be_{\perp }^x q^2 +
\frac{\varphi_\perp}{q^2}..) (1+ \sum_r \Omega_\perp^{(r)}(q^2))   \;,
\nonumber \\[0.1cm]
& H^{x}_\parallel &\;=\;&  (\al_{\parallel }^x + \be_{\parallel }^x q^2 +
\frac{\varphi_\parallel}{q^2} + .. ) (1+ \sum_r
\Omega_\parallel^{(r)}(q^2))
  \;,
\end{alignat}
where $\al$, $\be$ and $\varphi$  are complex numbers (c.f. again footnote
\ref{footnote:redundant}), the dots stand for higher orders in $q^2$ and
the functions 
$\Omega$ take into account resonance structures. Note there is no
$\varphi_0$-term since the photon does not have a zero helicity component.
The very same remarks, as  for the $\Delta$-functions in the previous
subsection, apply for the $\Omega$-terms. Depending on the region
it might be sufficient to approximate them with  a Breit-Wigner ansatz:
$\Omega_i^{(r)}(q^2) = \omega^{(\rho/\omega)}_i(q^2)/(q^2 -
m_{\rho/\omega}^2+ i m_{\rho/\omega} \Gamma_{\rho/\omega} ),
\omega^{(J/\Psi)}_i(q^2)/(q^2 - m_{J/\Psi}^2+ i m_{J/\Psi} \Gamma_{J/\Psi}
)$. Dependence on $q^2$ of
$\omega$ as well as polarisation dependence are a measure of
non-factorizable effects.
In fact  to some extent the residues $\omega^{(\rho/\omega)}$ and
$\omega^{(J/\Psi)}$ can be seen
as modelling the effect of a sum of resonances at the cost of the
$q^2$-dependence which is though not of major impact as long as one does
not get too close to the resonances.
By restricting the interval away from the $\rho$- and or
$J/\Psi$-resonances one could drop the $\Omega$-terms at the expense of
less statistics.
Note, 
the radiation of photons from light quarks, described at $q^2 =0$ by the
photon distribution amplitude, are mimicked at $q^2 > 1 \,\GeV^2$ by
$\langle \bar qq \rangle/q^2$-terms.
This leads to $1/q^4$-contributions in $H_{\parallel,\perp}$
\cite{Lyon:2013gba}
in processes such weak annihilation for example. These contributions are
though relatively small in the SM other than for the isospin asymmetry
(which is small by itself).

\subsection{Three momentum $\vv$ versus $q^2$}
\label{app:kq2}

In this appendix we give auxiliary formulae relating $\vv$ and $q^2$.
From \eqref{eq:speedD} one gets:
\begin{eqnarray}
\label{eq:k2q2}
\vv &=& \frac{1}{2 m_B} \sqrt{  (\epq - q^2)(\bar{q}^2_{\rm max}-q^2)} = \frac{1}{2 m_B}   \sqrt{\delta q^2 (\delta q^2 + 4 m_B m_{K^*})} \nonumber \\[0.1cm]
&=&  \sqrt{\delta q^2 ( m_{K^*}/m_B)}\left(1 + \frac{\delta q^2}{8 m_B m_{K^*}}   - 
\frac{(\delta q^2)^2}{16 m_B^2 m_{K^*}^2}    + .. \right)  
\;, \end{eqnarray}
where $\delta q^2$ is defined as
\begin{equation}
 \delta q^2 \equiv \epq-q^2 \;, \quad 
 \epq(\bar{q}^2_{\rm max}) = (m_B \mp m_{K^*})^2 \;,
\end{equation}
the (positive) distance to  the endpoint in the physical region. Thus to leading order $\vv^2 \propto 
\delta q^2$ or more precisely: 
\begin{equation}
 \delta q^2 = 2 m_B m_{K^*}(\sqrt{1 + \vv^2/m_{K^*}^2}-1) =\frac{m_B}{m_{K^*}} \vv^2 +{\cal{O}}(\vv^4) \; .
\end{equation}

\section{$F_L$ asymptotics}
\label{app:FLasymptotics}

For $B \to V \ell \ell$ the observable $F_L$  interpolates between
$F_L(0) = 0$ where the longitudinal mode decouples completely
and $F_L(\epq) = 1/3$ \eqref{eq:smaxlimitFL} where all polarisations are
equally probable.
This raises the question of whether anything can be said about  $F_L$ in
between
those two kinematic limits. The zero of the $A_{\rm FB}$ and the
equivalence
theorem
turn out to be of relevance in this context. For the subsequent analysis
it is helpful to rewrite
$F_L$ and $A_{\rm FB}$ in
the $\la = 0,\pm 1$ helicity and $V,A$ lepton-chirality basis:
\begin{equation}
\label{eq:genau}
A_{\rm FB} \propto {\rm Re}[ H_{-}^V H_{-}^{A*} -  H_+^V
H_+^{A*}  ]  \;, \quad F_L  = \frac{ |H_0^A|^2 + |H_0^V|^2}{ |H_0^A|^2 +
|H_0^V|^2 + |H_+^A|^2 + |H_+^V|^2 + |H_{-}^A|^2 + |H_{-}^V|^2 }
\;.
\end{equation}
The $V$-$A$ nature of the weak interaction distinguishes between the  
two $\pm$-helicity directions. In
particular for low $q^2$ this leads to $H^{V,A}_{+} \simeq 0$ which  
renders $H^{V,A}_{-}$ at the same time larger by roughly a factor of $
\sqrt{2}$. In the same kinematic region
the zero of $A_{\rm FB}(q_0^2)=0$ arises through $H^V_{-}\simeq 0$ as can
be seen from \eqref{eq:genau} which is due to  cancellations between
$C_9^{\rm eff}$ and the photon pole term described by $C_7^{\rm eff}$
in the lepton vector current. The LHCb collaboration has  
determined the location of
the zero $q_0^2= 4.9 \pm 0.9 \GeV^2$\cite{Aaij:2013iag} which is  
consistent with SM
predictions.
Naively, one might
expect $|H_0| : |H_{-}| : |H_{+}|  \simeq 1 : \sqrt{2} :0$ and
$F_L(q^2_0) \simeq (1^2+1^2)/(1^2+1^2 + \sqrt{2}^2)  = 1/2$.
To get a realistic number one needs to take into account the effect of
the prefactors of the polarisation vectors \eqref{eq:polis}, which are
related
to the equivalence theorem (relevant to heavy Higgs
boson physics for example {\it e.g.} \cite{DGH}).
The limit corresponds to $m_B^2 \gg m_{K^*}^2 , q^2$
with $q^2$ large enough such that the photon pole is not dominant.
Formally this happens as the $0$-helicity (longitudinal) polarisation
tensor
scales as $1/m_{K^*}$ and $1/\sqrt{q^2}$ \eqref{eq:polis} whereas the
other directions
do not encounter such an enhancement.  The zero helicity component can
therefore
be enhanced by a factor $\zeta \equiv (m_B^2/2)/ (m_{K^*} \sqrt{q^2})
|_{q^2 \simeq q_0^2} \simeq 7$ which leads to $F_L(q_0^2) \simeq   \zeta
(1^2+1^2) /( \zeta (1^2+1^2) + \sqrt{2}^2)
= 0.88$. This value is  close to the maximum of $F_L  
\simeq 0.8$ found in the literature, {\it e.g.} \cite{Bobeth:2010wg}.  
Differences can be understood by
the cancellation between  $C_7^{\rm eff}$ and $C_9^{\rm eff}$ terms in  
$|H_0^V|$.

In summary $F_L$ can be expected to raise considerably
from $1/3$ at the endpoint due to the equivalence theorem and then
asymptotes
to zero for low $q^2$ by virtue of the photon pole dominance. The effect
of the $zero$ of the $A_{\rm FB}$ enhances this effect and influences  
the maximum
of $F_L$.

\section{Tensor contributions to $d\Gamma/d\phi$ at the endpoint}
\label{app:rJ}

The quantity $r_\phi$ \eqref{eq:dGdphi}, in the dimension six basis, is given by \cite{Bobeth:2012vn}:
\begin{align}
r_\phi& =- \frac{4 }{9} \beta_\ell^2 \frac{|H_0|^2 +t_1}{(1+\frac{\beta_\ell^2}{3}) |H_0|^2+8 \frac{m_\ell^2}{q^2_{\rm max}} {\rm Re} (H_0^L H_0^{R*})+t_2} , \quad \quad  |H_0|^2 \equiv |H_0^L|^2+ |H_0^R|^2 \;,
\label{eq:rjcomplete}
\end{align}
where $|A_0| = |H_0|$ in the notation of \cite{Bobeth:2012vn}.
The symbols $t_{1,2}$ stand for tensor operator contributions, which read as
\begin{eqnarray}
t_1 &\equiv\;& -8 (|H_{t0}|^2+ |H_{+-}|^2) \;, \nonumber \\
 t_2  &\equiv\;& 4 (4-8/3 \beta_\ell^2)|H_{t0}|^2 +16/3 |H_{+-}|^2 +16 \sqrt{2} m_\ell /\sqrt{q^2_{\rm max}}
{\rm Re}((H_0^L+H_0^R) H_{t0}^*) \;. 
\end{eqnarray}
Here, we used
$|H_{+-}|=   | A_{\parallel \perp}|$,
$|H_{t0}|=   | A_{t0}|$ to translate from \cite{Bobeth:2012vn}.
Possible differences in phase conventions of polarisation vector should cancel as they always appear in the combination $\gpol(\la) \gpol(\la)^*$.

%\vspace*{-5mm}
%%%%%%%%%%%%%%%%%%%%%


\begin{thebibliography}{01}
\vspace*{3mm}

\bibitem{JW59}
  M.~Jacob and G.~C.~Wick,
  %``On the general theory of collisions for particles with spin,''
  Annals Phys.\  {\bf 7} (1959) 404
   [Annals Phys.\  {\bf 281} (2000) 774].
  %%CITATION = APNYA,7,404;%%
  %1054 citations counted in INSPIRE as of 02 Sep 2013

%\cite{Haber:1994pe}
\bibitem{Haber:1994pe}
  H.~E.~Haber,
  %``Spin formalism and applications to new physics searches,''
  In *Stanford 1993, Spin structure in high energy processes* 231-272
  [hep-ph/9405376].
  %%CITATION = HEP-PH/9405376;%%
  %54 citations counted in INSPIRE as of 02 Sep 2013


\bibitem{RZ13}
  R.~Zwicky,
  %``Endpoint symmetries of helicity amplitudes,''
  arXiv:1309.7802 [hep-ph].
  %%CITATION = ARXIV:1309.7802;%%
  
  
  
    %\cite{Grinstein:2004vb}
\bibitem{Grinstein:2004vb} 
  B.~Grinstein and D.~Pirjol,
  %``Exclusive rare B [*RIGHTWARDS ARROW*] K*** [*SCRIPT SMALL L*] + [*SCRIPT SMALL L*] - decays at low recoil: Controlling the long-distance effects,''
  Phys.\ Rev.\ D {\bf 70}, 114005 (2004)
  [hep-ph/0404250].
  %%CITATION = HEP-PH/0404250;%%
  
    %\cite{Grinstein:2002cz}
\bibitem{Grinstein:2002cz}
  B.~Grinstein and D.~Pirjol,
  %``Symmetry-breaking corrections to heavy meson form-factor relations,''
  Phys.\ Lett.\  B {\bf 533}, 8 (2002)
  [arXiv:hep-ph/0201298].
  %%CITATION = PHLTA,B533,8;%%

  
  %\cite{Isgur:1990kf}
\bibitem{Isgur:1990kf}
  N.~Isgur and M.~B.~Wise,
  %``RELATIONSHIP BETWEEN FORM-FACTORS IN SEMILEPTONIC anti-B AND D DECAYS AND EXCLUSIVE RARE anti-B MESON DECAYS,''
  Phys.\ Rev.\ D {\bf 42} (1990) 2388.
  %%CITATION = PHRVA,D42,2388;%%

%\cite{Bobeth:2010wg}
\bibitem{Bobeth:2010wg} 
  C.~Bobeth, G.~Hiller and D.~van Dyk,
  %``The Benefits of $\bar{B} -> \bar{K}^* l^+ l^-$ Decays at Low Recoil,''
  JHEP {\bf 1007}, 098 (2010)
  [arXiv:1006.5013 [hep-ph]].
  %%CITATION = ARXIV:1006.5013;%%
  
  
    %\cite{Bobeth:2012vn}
\bibitem{Bobeth:2012vn} 
  C.~Bobeth, G.~Hiller and D.~van Dyk,
  %``General Analysis of B -> K^(*) l^+ l^- Decays at Low Recoil,''
  Phys.\ Rev.\ D {\bf 87}, 034016 (2013),
    arXiv:1212.2321 [hep-ph].
  %%CITATION = ARXIV:1212.2321;%%



%\cite{Hambrock:2013zya}
\bibitem{Hambrock:2013zya} 
  C.~Hambrock, G.~Hiller, S.~Schacht and R.~Zwicky,
  %``B -> K^* Form Factors from Flavor Data to QCD and Back,''
  arXiv:1308.4379 [hep-ph].
  %%CITATION = ARXIV:1308.4379;%%

 
  
 \bibitem{PDG} 
  J.~Beringer {\it et al.}  [Particle Data Group Collaboration],
  %``Review of Particle Physics (RPP),''
  Phys.\ Rev.\ D {\bf 86}, 010001 (2012).
  %%CITATION = PHRVA,D86,010001;%%

\bibitem{HoZ13} 
  M.~Hopfer and R.~Zwicky, in preparation, Edinburgh 2014-xx.

    
    

  
  
\bibitem{Weinberg:1995mt}
  S.~Weinberg,
  ``The Quantum theory of fields. Vol. 1: Foundations,''
  Cambridge, UK: Univ. Pr. (1995) 609 p.

  
%\cite{Kruger:1999xa}
\bibitem{Kruger:1999xa} 
  F.~Kr\"uger, L.~M.~Sehgal, N.~Sinha and R.~Sinha,
  %``Angular distribution and CP asymmetries in the decays anti-B ---> K- pi+ e- e+ and anti-B ---> pi- pi+ e- e+,''
  Phys.\ Rev.\ D {\bf 61}, 114028 (2000)
  [Erratum-ibid.\ D {\bf 63}, 019901 (2001)]
  [hep-ph/9907386].
  %%CITATION = HEP-PH/9907386;%%
  %118 citations counted in INSPIRE as of 01 Sep 2013


  
  %\cite{Bobeth:2008ij}
\bibitem{Bobeth:2008ij} 
  C.~Bobeth, G.~Hiller and G.~Piranishvili,
  %``CP Asymmetries in bar $B \to \bar{K}^* (\to \bar{K} \pi) \bar{\ell} \ell$ and Untagged $\bar{B}_s$, $B_s \to \phi (\to K^{+} K^-) \bar{\ell} \ell$ Decays at NLO,''
  JHEP {\bf 0807}, 106 (2008)
  [arXiv:0805.2525 [hep-ph]].
  %%CITATION = ARXIV:0805.2525;%%
  %129 citations counted in INSPIRE as of 02 Sep 2013
  
    

    
%\cite{Altmannshofer:2008dz}
\bibitem{Altmannshofer:2008dz}
  W.~Altmannshofer {\it et al.},
  %``Symmetries and Asymmetries of $B \to K^{*} \mu^{+} \mu^{-}$ Decays in the
  %Standard Model and Beyond,''
  JHEP {\bf 0901} (2009) 019
  [arXiv:0811.1214 [hep-ph]].
  %%CITATION = JHEPA,0901,019;%%
  

  %\cite{DescotesGenon:2012zf}
\bibitem{DescotesGenon:2012zf} 
  S.~Descotes-Genon, J.~Matias, M.~Ramon and J.~Virto,
  %``Implications from clean observables for the binned analysis of $B -> K*\mu^+\mu^-$ at large recoil,''
  JHEP {\bf 1301}, 048 (2013)
  [arXiv:1207.2753 [hep-ph]].
  %%CITATION = ARXIV:1207.2753;%%

     %\cite{Kruger:2005ep}
\bibitem{Kruger:2005ep} 
  F.~Kr\"uger and J.~Matias,
  %``Probing new physics via the transverse amplitudes of B0 ---> K*0 (---> K- pi+) l+l- at large recoil,''
  Phys.\ Rev.\ D {\bf 71}, 094009 (2005)
  [hep-ph/0502060].
  %%CITATION = HEP-PH/0502060;%%
  
     %\cite{Hambrock:2012dg}
\bibitem{Hambrock:2012dg}
 C.~Hambrock and G.~Hiller,
 %``Extracting $B \to K^*$ Form Factors from Data,''
 Phys.\ Rev.\ Lett.\  {\bf 109}, 091802 (2012)
 [arXiv:1204.4444 [hep-ph]].
 %%CITATION = ARXIV:1204.4444;%%



\bibitem{BaBarLakeLouise}
S.~Akar [BaBar Collaboration], Lake Louise Winter Institute, Canada, February 23, 2012.

\bibitem{HidekiICHEP2012}
  CDF note 10894, July 2012,
%H. Miyake,  at
%the 36th International Conference for High Energy Physics (ICHEP), July 4-11, 2012, Melbourne, %Australia.
http://www-cdf.fnal.gov/physics/new/bottom/.
%120628.blessed-b2smumu_96/public_b2smumu.pdf.
  
%\cite{Aaij:2013iag}
\bibitem{Aaij:2013iag} 
  R.~Aaij {\it et al.}  [LHCb Collaboration],
  %``Differential branching fraction and angular analysis of the decay $B^{0} \to K^{*0} \mu^{+}\mu^{-}$,''
  arXiv:1304.6325 [hep-ex].
  %%CITATION = ARXIV:1304.6325;%%


  %\cite{Aaij:2013aln}
\bibitem{Aaij:2013aln} 
  R.~Aaij {\it et al.}  [ LHCb Collaboration],
  %``Differential branching fraction and angular analysis of the decay $B_s^0\to\phi\mu^{+}\mu^{-}$,''
  JHEP {\bf 1307}, 084 (2013)
  [arXiv:1305.2168 [hep-ex]].
  %%CITATION = ARXIV:1305.2168;%%



%\cite{Aaij:2013qta}
\bibitem{Aaij:2013qta} 
  R.~Aaij {\it et al.}  [LHCb Collaboration],
  %``Measurement of form-factor independent observables in the decay $B^{0} \to K^{*0} \mu^+ \mu^-$,''
  arXiv:1308.1707 [hep-ex].
  %%CITATION = ARXIV:1308.1707;%%




%\cite{ATLAS:2013ola}
\bibitem{ATLAS:2013ola} 
  [ATLAS Collaboration],
  %``Angular Analysis of $B_{d} \to K^{\ast 0}\mu^{+}\mu^{-}$ with the ATLAS Experiment,''
  ATLAS-CONF-2013-038.
  %%CITATION = ATLAS-CONF-2013-038;%%

%\cite{CMS:cwa}
\bibitem{CMS:cwa} 
  S.~Chatrchyan {\it et al.}  [ CMS Collaboration],
  %``Angular analysis and branching fraction measurement of the decay B0 to K*0 mu+ mu-,''
  arXiv:1308.3409 [hep-ex].
  %%CITATION = ARXIV:1308.3409;%%
  
  
  
  
  %\cite{Altmannshofer:2009ma}
\bibitem{Altmannshofer:2009ma} 
  W.~Altmannshofer, A.~J.~Buras, D.~M.~Straub and M.~Wick,
  %``New strategies for New Physics search in $B \to K^{*} \nu \bar{\nu}$, $B \to K \nu \bar{\nu}$ and $B \to X_{s} \nu \bar{\nu}$ decays,''
  JHEP {\bf 0904}, 022 (2009)
  [arXiv:0902.0160 [hep-ph]].
  %%CITATION = ARXIV:0902.0160;%%
  %75 citations counted in INSPIRE as of 01 Sep 2013
  
  
      %\cite{Ligeti:1995yz}
\bibitem{Ligeti:1995yz} 
  Z.~Ligeti and M.~B.~Wise,
  %``|V(ub)| from exclusive B and D decays,''
  Phys.\ Rev.\ D {\bf 53}, 4937 (1996)
  [hep-ph/9512225].
  %%CITATION = HEP-PH/9512225;%%
  
  %\cite{Kruger:1996cv}
\bibitem{Kruger:1996cv} 
  F.~Kruger and L.~M.~Sehgal,
  %``Lepton polarization in the decays b ---> X(s) mu+ mu- and B ---> X(s) tau+ tau-,''
  Phys.\ Lett.\ B {\bf 380}, 199 (1996)
  [hep-ph/9603237].
  %%CITATION = HEP-PH/9603237;%%
  
  %\cite{Ali:1999mm}
\bibitem{Ali:1999mm} 
  A.~Ali, P.~Ball, L.~T.~Handoko and G.~Hiller,
  %``A Comparative study of the decays $B \to$ ($K$, $K^{*)} \ell^+ \ell^-$ in standard model and supersymmetric theories,''
  Phys.\ Rev.\ D {\bf 61}, 074024 (2000)
  [hep-ph/9910221].
  %%CITATION = HEP-PH/9910221;%%



  %\cite{Aaij:2013pta}
\bibitem{Aaij:2013pta} 
  R.~Aaij {\it et al.}  [LHCb Collaboration],
  %``Observation of a resonance in $B^+ \to K^+ \mu^+\mu^-$ decays at low recoil,''
  arXiv:1307.7595 [hep-ex].
  %%CITATION = ARXIV:1307.7595;%%
  
  
  
  %\cite{Bobeth:2011gi}
\bibitem{Bobeth:2011gi} 
  C.~Bobeth, G.~Hiller and D.~van Dyk,
  %``More Benefits of Semileptonic Rare B Decays at Low Recoil: CP Violation,''
  JHEP {\bf 1107}, 067 (2011)
  [arXiv:1105.0376 [hep-ph]].
  %%CITATION = ARXIV:1105.0376;%%

  
    %\cite{Dimou:2012un}
\bibitem{Dimou:2012un}
  M.~Dimou, J.~Lyon and R.~Zwicky,
  %``Exclusive Chromomagnetism in heavy-to-light FCNCs,''
  Phys.\ Rev.\ D {\bf 87} (2013) 074008
  [arXiv:1212.2242 [hep-ph]].
  %%CITATION = ARXIV:1212.2242;%%
  %6 citations counted in INSPIRE as of 23 Oct 2013
   
   
     %\cite{Ball:2004rg}
\bibitem{Ball:2004rg}
  P.~Ball and R.~Zwicky,
  %``B(D,S) ---> rho, omega, K*, phi decay form-factors from light-cone sum rules revisited,''
  Phys.\ Rev.\ D {\bf 71} (2005) 014029
  [hep-ph/0412079].
  %%CITATION = HEP-PH/0412079;%%
  %368 citations counted in INSPIRE as of 23 Oct 2013
  
  %\cite{Horgan:2013hoa}
\bibitem{Horgan:2013hoa}
  R.~R.~Horgan, Z.~Liu, S.~Meinel and M.~Wingate,
  %``Lattice QCD calculation of form factors describing the rare decays B to K^* l^+ l^- and B_s to phi l^+ l^-,''
  arXiv:1310.3722 [hep-lat].
  %%CITATION = ARXIV:1310.3722;%%
  %2 citations counted in INSPIRE as of 23 Oct 2013

   
    \bibitem{Ball:2006eu}
  P.~Ball, G.~W.~Jones and R.~Zwicky,
%  ``B ---> V gamma beyond QCD factorisation,''
  Phys.\ Rev.\ D {\bf 75} (2007) 054004
  [hep-ph/0612081].
  %%CITATION = HEP-PH/0612081;%%
  %145 citations counted in INSPIRE as of 03 Sep 2013
  
  
  %\cite{Aaij:2013oba}
\bibitem{Aaij:2013oba}
  R.~Aaij {\it et al.}  [LHCb Collaboration],
  %``Measurement of $CP$ violation and the $B_s^0$ meson decay width difference with $B_s^0 \to J/\psi K^+K^-$ and $B_s^0\to J/\psi\pi^+\pi^-$ decays,''
  Phys.\ Rev.\ D {\bf 87} (2013) 112010
  [arXiv:1304.2600 [hep-ex]].
  %%CITATION = ARXIV:1304.2600;%%
  %33 citations counted in INSPIRE as of 25 Oct 2013
  
  %\cite{Aubert:2007hz}
\bibitem{Aubert:2007hz} 
  B.~Aubert {\it et al.}  [BaBar Collaboration],
  %``Measurement of decay amplitudes of $B \to J/\psi K^{*}, \psi(2S) K^{*}$, and $\chi_{c1} K^{*}$ with an angular analysis,''
  Phys.\ Rev.\ D {\bf 76}, 031102 (2007)
  [arXiv:0704.0522 [hep-ex]].
  %%CITATION = ARXIV:0704.0522;%%
  
 %\cite{Aaij:2013cma}
\bibitem{Aaij:2013cma}
  RAaij {\it et al.}  [LHCb Collaboration],
  %``Measurement of the polarization amplitudes  in $B^0 \to J/\psi K^{*}(892)^0$ decays,''
  Phys.\ Rev.\ D {\bf 88} (2013) 052002
  [arXiv:1307.2782 [hep-ex]].
  %%CITATION = ARXIV:1307.2782;%%
  %1 citations counted in INSPIRE as of 25 Nov 2013
  
  %\cite{Beylich:2011aq}
\bibitem{Beylich:2011aq} 
  M.~Beylich, G.~Buchalla and T.~Feldmann,
  %``Theory of B -> K(*)l+l- decays at high q^2: OPE and quark-hadron duality,''
  Eur.\ Phys.\ J.\ C {\bf 71}, 1635 (2011)
  [arXiv:1101.5118 [hep-ph]].
  %%CITATION = ARXIV:1101.5118;%%

     
 % \cite{Bobeth:2007dw}
\bibitem{Bobeth:2007dw}
  C.~Bobeth, G.~Hiller and G.~Piranishvili,
  %``Angular distributions of anti-B ---> K anti-l l decays,''
  JHEP {\bf 0712} (2007) 040
  [arXiv:0709.4174 [hep-ph]].
  %%CITATION = ARXIV:0709.4174;%%
  %69 citations counted in INSPIRE as of 05 Oct 2013
  
  
    %\cite{Becirevic:2012dp}
\bibitem{Becirevic:2012dp} 
  D.~Becirevic and A.~Tayduganov,
  %``Impact of $B\to K^\ast_0 \ell^+\ell^-$ on the New Physics search in $B\to K^\ast \ell^+\ell^-$ decay,''
  Nucl.\ Phys.\ B {\bf 868}, 368 (2013)
  [arXiv:1207.4004 [hep-ph]].
  %%CITATION = ARXIV:1207.4004;%%
  %19 citations counted in INSPIRE as of 12 Sep 2013
  
  
  %\cite{Lyon:2013gba}
\bibitem{Lyon:2013gba}
  J.~Lyon and R.~Zwicky,
  %``Isospin asymmetries in $B\to (K^*,\rho) \gamma/ l^+ l^-$ and $B \to K l^+ l^-$ in and beyond the Standard Model,''
  arXiv:1305.4797 [hep-ph],  to appear in PRD
  %%CITATION = ARXIV:1305.4797;%%
  %7 citations counted in INSPIRE as of 29 Sep 2013

    
  %\cite{Li:2010ra}
\bibitem{Li:2010ra}
  R.~-H.~Li, C.~-D.~Lu and W.~Wang,
  %``Branching ratios, forward-backward asymmetries and angular distributions of $B\to K_2^*l^+l^-$ in the standard model and new physics scenarios,''
  Phys.\ Rev.\ D {\bf 83} (2011) 034034
  [arXiv:1012.2129 [hep-ph]].
  %%CITATION = ARXIV:1012.2129;%%
  %10 citations counted in INSPIRE as of 05 Oct 2013
  
  
  
    %%%% dibaryons
  
    
    %\cite{Aubert:2009qz}
\bibitem{Aubert:2009qz} 
  B.~Aubert {\it et al.}  [BaBar Collaboration],
  %``Observation and study of baryonic B decays: B ---> D(*) p anti-p, D(*) p anti-p pi, and D(*) p anti-p pi pi,''
  arXiv:0908.2202 [hep-ex].
  %%CITATION = ARXIV:0908.2202;%%
  %2 citations counted in INSPIRE as of 05 Oct 2013
  

  
  %\cite{Chang:2008yw}
\bibitem{Chang:2008yw} 
  Y.~-W.~Chang {\it et al.}  [Belle Collaboration],
  %``Observation of B0 ---> Lambda anti-Lambda K0 and B0 to Lambda anti-Lambda K*0 at Belle,''
  Phys.\ Rev.\ D {\bf 79}, 052006 (2009)
  [arXiv:0811.3826 [hep-ex]].
  %%CITATION = ARXIV:0811.3826;%%
  %9 citations counted in INSPIRE as of 05 Oct 2013
  
%\cite{Chen:2008jy}
\bibitem{Chen:2008jy} 
  J.~H.~Chen {\it et al.}  [Belle Collaboration],
  %``Observation of B0 ---> p anti-p K*0 with a large K*0 polarization,''
  Phys.\ Rev.\ Lett.\  {\bf 100}, 251801 (2008)
  [arXiv:0802.0336 [hep-ex]].
  %%CITATION = ARXIV:0802.0336;%%
  %19 citations counted in INSPIRE as of 05 Oct 2013
  

  
  %\cite{Aaij:2013yba}
\bibitem{Aaij:2013yba} 
  R.~Aaij {\it et al.}  [LHCb Collaboration],
  %``Searches for $B^0_{(s)} \to J/\psi p\bar{p}$ and $B^+ \to J/\psi p\bar{p}\pi^+$ decays,''
  JHEP {\bf 1309}, 006 (2013)
  [arXiv:1306.4489 [hep-ex]].
  %%CITATION = ARXIV:1306.4489;%%
  
    %\cite{Kagan:2004uw}
\bibitem{Kagan:2004uw}
  A.~L.~Kagan,
  %``Polarization in B ---> VV decays,''
  Phys.\ Lett.\ B {\bf 601} (2004) 151
  [hep-ph/0405134].
  %%CITATION = HEP-PH/0405134;%%
  %181 citations counted in INSPIRE as of 14 Nov 2013
    
 
 %
\bibitem{Beneke:2006hg}
  M.~Beneke, J.~Rohrer and D.~Yang,
  %``Branching fractions, polarisation and asymmetries of B ---> VV decays,''
  Nucl.\ Phys.\ B {\bf 774} (2007) 64
  [hep-ph/0612290].
  %%CITATION = HEP-PH/0612290;%%
  %143 citations counted in INSPIRE as of 25 Nov 2013
 
 
 %\cite{Aaij:2013qha}
\bibitem{Aaij:2013qha}
  RAaij {\it et al.}  [LHCb Collaboration],
  %``First measurement of the CP-violating phase in $B_s^0 \to \phi \phi$ decays,''
  Phys.\ Rev.\ Lett.\  {\bf 110} (2013) 241802
  [arXiv:1303.7125 [hep-ex]].
  %%CITATION = ARXIV:1303.7125;%%
  %10 citations counted in INSPIRE as of 25 Nov 2013
 
 %\cite{Aaij:2013gga}
\bibitem{Aaij:2013gga}
  RAaij {\it et al.}  [LHCb Collaboration],
  %``First observation of the decay $B_s^0 \rightarrow \phi \bar{K}^{*0}$,''
  JHEP {\bf 1311} (2013) 092
  [arXiv:1306.2239 [hep-ex]].
  %%CITATION = ARXIV:1306.2239;%%
  %2 citations counted in INSPIRE as of 25 Nov 2013
 
    
    %\cite{Matias:2012qz}
\bibitem{Matias:2012qz} 
  J.~Matias,
  %``On the S-wave pollution of B-> K* l+l- observables,''
  Phys.\ Rev.\ D {\bf 86}, 094024 (2012)
  [arXiv:1209.1525 [hep-ph]].
  %%CITATION = ARXIV:1209.1525;%%
  %16 citations counted in INSPIRE as of 12 Sep 2013
  

  
  %\cite{Blake:2012mb}
\bibitem{Blake:2012mb} 
  T.~Blake, U.~Egede and A.~Shires,
  %``The effect of S-wave interference on the $B^0 \to K^{\ast 0}\ell^+\ell^-$ angular observables,''
  JHEP {\bf 1303}, 027 (2013)
  [arXiv:1210.5279 [hep-ph]].
  %%CITATION = ARXIV:1210.5279;%%
  %10 citations counted in INSPIRE as of 12 Sep 2013


    %\cite{Descotes-Genon:2013wba}
\bibitem{Descotes-Genon:2013wba} 
  S.~Descotes-Genon, J.~Matias and J.~Virto,
  %``Understanding the B->K*mu+mu- Anomaly,''
  Phys.\  Rev.\  D 88, {\bf 074002} (2013)
  [arXiv:1307.5683 [hep-ph]].
  %%CITATION = ARXIV:1307.5683;%%
  %13 citations counted in INSPIRE as of 21 Oct 2013
  
  
  %\cite{Altmannshofer:2013foa}
\bibitem{Altmannshofer:2013foa} 
  W.~Altmannshofer and D.~M.~Straub,
  %``New physics in B -> K*{\mu}{\mu}?,''
  arXiv:1308.1501 [hep-ph].
  %%CITATION = ARXIV:1308.1501;%%
  %10 citations counted in INSPIRE as of 21 Oct 2013


%\cite{Beaujean:2013soa}
\bibitem{Beaujean:2013soa} 
  F.~Beaujean, C.~Bobeth and D.~van Dyk,
  %``Comprehensive Bayesian Analysis of Rare (Semi)leptonic and Radiative B Decays,''
  arXiv:1310.2478 [hep-ph].
  %%CITATION = ARXIV:1310.2478;%%
  %2 citations counted in INSPIRE as of 21 Oct 2013

  %\cite{Bobeth:2011nj}
\bibitem{Bobeth:2011nj} 
  C.~Bobeth, G.~Hiller, D.~van Dyk and C.~Wacker,
  %``The Decay B --> K l^+ l^- at Low Hadronic Recoil and Model-Independent Delta B = 1 Constraints,''
  JHEP {\bf 1201}, 107 (2012)
  [arXiv:1111.2558 [hep-ph]].
  %%CITATION = ARXIV:1111.2558;%%
  %57 citations counted in INSPIRE as of 07 Oct 2013
  
  %\cite{Egede:2010zc}
\bibitem{Egede:2010zc}
  U.~Egede, T.~Hurth, J.~Matias, M.~Ramon and W.~Reece,
  %``New physics reach of the decay mode $\bar{B} \to \bar{K}^{*0}\ell^+\ell^-$,''
  JHEP {\bf 1010} (2010) 056
  [arXiv:1005.0571 [hep-ph]].
  %%CITATION = ARXIV:1005.0571;%%
  %66 citations counted in INSPIRE as of 28 Oct 2013
  
  \bibitem{Reece}
Reece, W. R. (2010). PhD thesis,  Imperial College London.
http://cds.cern.ch/record/1280731


  \bibitem{Petridis} K.~Petridis, talk given at the UK flavour workshop, Durham, United Kingdom, September 4-7, 2013.

 \bibitem{DGH}
  J.~F.~Donoghue, E.~Golowich and B.~R.~Holstein,
  %``Dynamics of the standard model,''
  Camb.\ Monogr.\ Part.\ Phys.\ Nucl.\ Phys.\ Cosmol.\  {\bf 2} (1992) 1.
  %%CITATION = CMPCE,2,1;%%

\end{thebibliography}
\end{document}